\documentclass[10pt,twocolumn,secnumarabic, nobibnotes, aps, prd]{revtex4-2}
\usepackage{eurosym}
\usepackage{xcolor,soul}
\usepackage{amsmath}
\usepackage{wrapfig}
\usepackage[utf8]{inputenc}
\usepackage{lipsum}
\usepackage{subfigure}
\usepackage{xparse,xcoffins}
\usepackage{graphicx}
\usepackage{calligra}

\DeclareMathAlphabet{\mathcalligra}{T1}{calligra}{m}{n}
\DeclareFontShape{T1}{calligra}{m}{n}{<->s*[2.2]callig15}{}

\sethlcolor{red}

\ExplSyntaxOn
\NewCoffin\imagecoffin
\NewCoffin\labelcoffin
\keys_define:nn { miguel/label }
 {
  label   .tl_set:N = \l_miguel_label_tl,
  labelbox .bool_set:N = \l_miguel_label_box_bool,
  labelbox .default:n = true,
  fontsize .tl_set:N = \l_miguel_label_size_tl,
  fontsize .initial:n = \footnotesize,
  pos .choice:,
  pos/nw .code:n = \tl_set:Nn \l_miguel_label_pos_tl { left,up },
  pos/ne .code:n = \tl_set:Nn \l_miguel_label_pos_tl { right,up },
  pos/sw .code:n = \tl_set:Nn \l_miguel_label_pos_tl { left,down },
  pos/se .code:n = \tl_set:Nn \l_miguel_label_pos_tl { right,down },
  pos/n .code:n = \tl_set:Nn \l_miguel_label_pos_tl { hc,up },
  pos/w .code:n = \tl_set:Nn \l_miguel_label_pos_tl { left,vc },
  pos/s .code:n = \tl_set:Nn \l_miguel_label_pos_tl { hc,down },
  pos/e .code:n = \tl_set:Nn \l_miguel_label_pos_tl { right,vc },
  pos .initial:n = nw,
  unknown .code:n   = \clist_put_right:Nx \l_miguel_label_clist
                       { \l_keys_key_tl = \exp_not:n { #1 } }
 }
\clist_new:N \l_miguel_label_clist
\box_new:N \l_miguel_label_box
\box_new:N \l_miguel_label_image_box
\NewDocumentCommand{\xincludegraphics}{O{}m}
 {
  \group_begin:
  \tl_clear:N \l_miguel_label_tl
  \clist_clear:N \l_miguel_label_clist
  \keys_set:nn { miguel/label } { #1 }
  \tl_if_empty:NTF \l_miguel_label_tl
   {
    \miguel_includegraphics:Vn \l_miguel_label_clist { #2 }
   }
   {
    \SetHorizontalCoffin\imagecoffin
     {
      \miguel_includegraphics:Vn \l_miguel_label_clist { #2 }
     }
    \SetHorizontalCoffin\labelcoffin
     {
      \raisebox{\depth}
       {
        \bool_if:NTF \l_miguel_label_box_bool
         { \fcolorbox{white}{white}{\l_miguel_label_size_tl\l_miguel_label_tl} }
         { \l_miguel_label_size_tl\l_miguel_label_tl }
       }
     }
    \SetVerticalPole\imagecoffin{left}{3pt+\CoffinWidth\labelcoffin/2}
    \SetVerticalPole\imagecoffin{right}{\Width-3pt-\CoffinWidth\labelcoffin/2}
    \SetHorizontalPole\imagecoffin{up}{\Height-3pt-\CoffinHeight\labelcoffin/2}
    \SetHorizontalPole\imagecoffin{down}{3pt+\CoffinHeight\labelcoffin/2}
    \use:x{\JoinCoffins\imagecoffin[\l_miguel_label_pos_tl]\labelcoffin[vc,hc]}
    \TypesetCoffin\imagecoffin
   }
   \group_end:
 }
\NewDocumentCommand{\setlabel}{m}
 {
  \keys_set:nn { miguel/label } { #1 }
 }
\cs_new_protected:Nn \miguel_includegraphics:nn
 {
  \includegraphics[#1]{#2}
 }
\cs_generate_variant:Nn \miguel_includegraphics:nn { V }
\ExplSyntaxOff
\makeatletter
\newcommand{\twocolumncaption}{\@dblarg\@twocolumncaption}
\def\@twocolumncaption[#1]#2{  \renewcommand{\@makecaption}[2]{    \par\vskip\abovecaptionskip\begingroup\small\rmfamily
    \splittopskip=0pt
    \setbox\@tempboxa=\vbox{
      \@arrayparboxrestore \let \\\@normalcr
      \hsize=.5\hsize \advance\hsize-1em
      \let\\\heading@cr
      \noindent ##1\ ##2\par    }    \vbadness=10000
    \setbox\z@=\vsplit\@tempboxa to .55\ht\@tempboxa
    \setbox\z@=\vtop{\hrule height 0pt \unvbox\z@}
    \setbox\tw@=\vtop{\hrule height 0pt \unvbox\@tempboxa}
    \noindent\box\z@\hfill\box\tw@\par
    \endgroup\vskip \belowcaptionskip
  }  \setlength{\abovecaptionskip}{4ex}  \caption[#1]{#2}}

\begin{document}

\title{Toroidal Confinement and Beyond: Vorticity-Defined Morphologies of
Dipolar $^{164}$Dy Quantum Droplets}
\author{S. Sanjay$^1$}
\email{s\textunderscore sanjay@cb.students.amrita.edu}
\author{S. Saravana Veni$^1$}
\email{s\textunderscore saravanaveni@cb.amrita.edu.in}
\author{Boris A. Malomed$^{2,3}$}
\email{malomed@tauex.tau.ac.il}
\affiliation{$^1$Department of Physics, Amrita School of Physical Sciences,
Amrita Vishwa Vidyapeetham, Coimbatore-641112, Tamil Nadu, India.}
\affiliation{$^2$Department of Physical Electronics, School of Electrical Engineering, Faculty of Engineering,
Tel Aviv University, Tel Aviv 69978, Israel.}
\affiliation{$^3$Instituto de Alta Investigaci\'{o}n, Universidad de
Tarapac\'{a}, Casilla 7D, Arica, Chile.}

\begin{abstract}
We investigate the formation, stability, and dynamics of 3D ring-shaped and
multipole vortical quantum droplets (QDs) in non-rotating dipolar
Bose-Einstein condensates held in a toroidal trapping potential. The QD
dynamics are investigated in the framework of the extended Gross-Pitaevskii
equation, which includes long-range dipole-dipole interactions (DDI) and the
beyond-mean-field Lee-Huang-Yang (LHY) term, revealing the emergence of
self-bound states. Stable stationary solutions for multipole QDs with
different values of the topological charge (vorticity $S$) are shaped as
necklace-like modes, with the number of \textquotedblleft beads"
(multipole's order) $n=2S$, up to $S=6$. The stability area of the
multipoles shrinks with the increase of $S$. For higher values of $S$ the centrifugal effect associated with the phase winding destabilizes the annular density and drives the formation of fragmented multipole droplet states.
The dependence of the chemical potential, total energy and peak density
on the norm (number of particles)
and $S$ is produced. These findings uncover the stabilizing effect of the
LHY correction and DDI anisotropy in maintaining complex QD states in the
non-rotating configurations.
\end{abstract}

\maketitle


\section{Introduction}

A great advancement has been achieved in recent studies of macroscopic
quantum phenomena such as formation of quantum droplets (QDs) \cite%
{Bulgac,luo2021new, guo2021new}, supersolidity \cite%
{boninsegni2012colloquium,recati2023supersolidity,ilzhofer2021phase},
vortices \cite{verhelst2017vortex, klaus2022observation, chaika2023making},
and others. Specifically, the investigation of QDs, which are sustained by
the stable balance between the mean-field (MF) interactions and beyond-MF
[Lee-Huang-Yang (LHY)] corrections to them, induced by quantum fluctuations
\cite{petrov,Astrakharchik}, offers a broad framework for the development of
these studies, owing to\ the relative simplicity of the theoretical model
and their availability for the experiment \cite%
{neely2010observation,weiler2008spontaneous}. The so predicted QD stability
within the MF-LHY framework aligns well with outcomes generated by the Monte
Carlo method applied to the corresponding many-body setting \cite%
{macia2016droplets,cinti2017superfluid,cikojevic2018ultradilute,cikojevic2019universality,cikojevic2020finite,parisi2019liquid}%
.

In the experiment, QDs were initially created in Bose-Einstein condensates
(BECs) of magnetic atoms \cite{ferrier2016observation,bottcher2020new}. A
great deal of interest has also been drawn to QDs, filled by ultra-diluted
superfluids, in homonuclear \cite{cheiney2018bright,semeghini2018self} and
heteronuclear \cite{d2019observation,burchianti2020dual} binary BECs with
contact inter-atomic interactions. In the latter case, the formation of
stable QDs is provided by the possibility to properly adjust the
corresponding atomic collision length by dint of the Feshbach-resonance
effect, imposed by the DC magnetic field \cite{d2007feshbach}. In addition
to the formation of QDs, also studied were collisions between moving QDs in
one-dimensional (1D) \cite{katsimiga2023interactions}, 2D \cite%
{hu2022collisional}, and 3D \cite{ferioli2019collisions} geometries, along
with the experimentally observed effectively 1D scattering of QDs on
localized potentials \cite{debnath2023interaction}. On the other hand, the
long-range dipole-dipole interactions (DDI) in the BEC of magnetic atoms
enable the realization of stable anisotropic QD configurations \cite%
{Pfau1,Pfau2,wachtler2016ground,xi2016droplet,Boronat}. In particular, QD
structures in 3D harmonic-oscillator (HO) and potential-box potentials have
been studied \cite{young2022supersolid,young2022supersolidbox}. Ground-state
and metastable striped patterns (superstripe states), trapped in the 3D HO
potentials, were reported too \cite{young2023mini}. Beyond the BEC realm,
studies of droplet phenomena have been extended to vapors \cite%
{holyst2013evaporation} and nonlinear photonic systems \cite%
{wilson2018observation,s_sanjay,deeskhita}.

An obviously interesting but challenging possibility is the incorporation of
vorticity into self-bound QDs. The self-interaction causes azimuthal
instability, which tends to split the 3D vortex into fragments \cite%
{Mihalache1,book,Mihalache2}. This issue can also be resolved with the help
of the LHY corrections. The balance between the self-repulsive LHY terms and
the MF attractive ones results in stable configurations, which is similar to
the mechanism known in nonlinear optics, where the stability may be provided
by the competitive cubic-quintic nonlinearity \cite%
{mihalache2002stable,reyna2017high}. Thus, stable 2D \cite%
{li2018two,dong2022internal,dong2022bistable} and 3D \cite%
{kartashov2018three} vortex QDs (VQDs) with the embedded angular momentum
have been predicted, as summarized in Ref. \cite{Li2024CanVQ}. Metastable
ring-shaped QD clusters and rotating VQD clusters carrying multiple
vorticity were predicted too \cite%
{kartashov2019metastability,tengstrand2019rotating}. The VQD stability can
be enhanced by trapping the BEC in a toroidal potential \cite{LDong1}.
Toroidal traps are especially adept at accommodating vortex states due to
their inherent rotational symmetry and ability to sustain persistent
currents \cite{persistent-currents}. In particular, the formation of 3D
stable vortices and multipole QDs (necklace-shaped chains) under the
toroidal confinement were demonstrated up to vorticity of $S=12$ \cite%
{LDong1}. In the absence of the trapping potential, VQDs with high
topological charges require an enormous number of atoms to retain their
stability \cite{kartashov2018three,li2018two}. A similar mechanism ensures
the stabilization of 2D VQDs with different vorticities \cite{LDong2,LDong3}
and multipole 2D QDs \cite{LDong4} by means of the annular trap.

VQDs in dipolar BECs were studied too \cite{cidrim2018vortices}, where the
long-range DDI is fundamentally different from the contact interactions in
binary condensates. These studies further demonstrate that isotropic vortex
solutions, with dipoles polarized parallel to the vortical axis, are
inherently unstable. Stable VQDs have been predicted in 2D anisotropic
dipolar configurations, with the dipoles polarized at an angle to the pivot
\cite{li2024two}. Despite the significant progress in the studies of VQDs,
multipole QDs in dipolar BEC trapped in the 3D toroidal geometries have not
been addressed as yet, which is the objective of the present work.
Experimentally, toroidal potentials have been employed to probe superfluid
properties, such as the quantized circulation and stability thresholds of
long-lived states~\cite{murray2013probing,wright2013threshold}. These
ring-shaped traps support topologically nontrivial structures, including
persistent currents~\cite{munoz2015persistent,tengstrand2021persistent} and
knotted vortex lines. They have also been proposed as platforms for
exploring entanglement phenomena in expanding BECs~\cite%
{bhardwaj2024entanglement}. Recent work~\cite{bazhan2022generation} has
predicted the formation of Josephson vortices in a stacked pair of toroidal
BECs, in the framework of a dissipative Gross-Pitaevskii (GP) equations.

Motivated by the fact that the toroidal potentials, when combined with the
repulsive LHY nonlinearity, enhances the robustness of the self-bound QDs, where the
toroidal trap creates the localized-density states, and the self-repulsive LHY helps in
stabilizing them. In the droplets carrying vorticity $S$, the centrifugal contribution
to the kinetic energy, generated by the phase winding, which is proportional to $S^{2}$,
pushes the density outward and competes with the toroidal confinement and LHY
repulsion. The competition leads to the spontaneous emergence of
multipole droplet structures, which is the central result
of the present work. The present work aims to systematically investigate structural
characteristics and stability of ring-shaped and multipole QDs in the 3D
dipolar BEC confined by the toroidal potential. Our numerical analysis
combines the imaginary-time propagation (ITP) and real-time propagation
methods, to produce both stationary and evolving solutions, focusing on
their physically relevant characteristics, such as the chemical potential,
peak density, radial and axial widths, and vortex-core structure, for
different winding numbers (vorticity values).
The paper is organized as follows: in Section~\ref{model}, we present the
model for the dipolar BEC, which includes the MF and LHY interactions, along
with the toroidal confinement. Section~\ref{results} is divided in three
parts: \ref{stat_ste} addresses stationary states of multipole droplets
under toroidal confinement, in III.2 
we examine their stability under perturbations, and in III.3 
we present the stationary states
under the action of the Gaussian confinement. The paper is concluded by
Section IV.
\begin{figure}[tbp]
\centering
\xincludegraphics[width=0.45\linewidth, label={$\qquad$a)}]{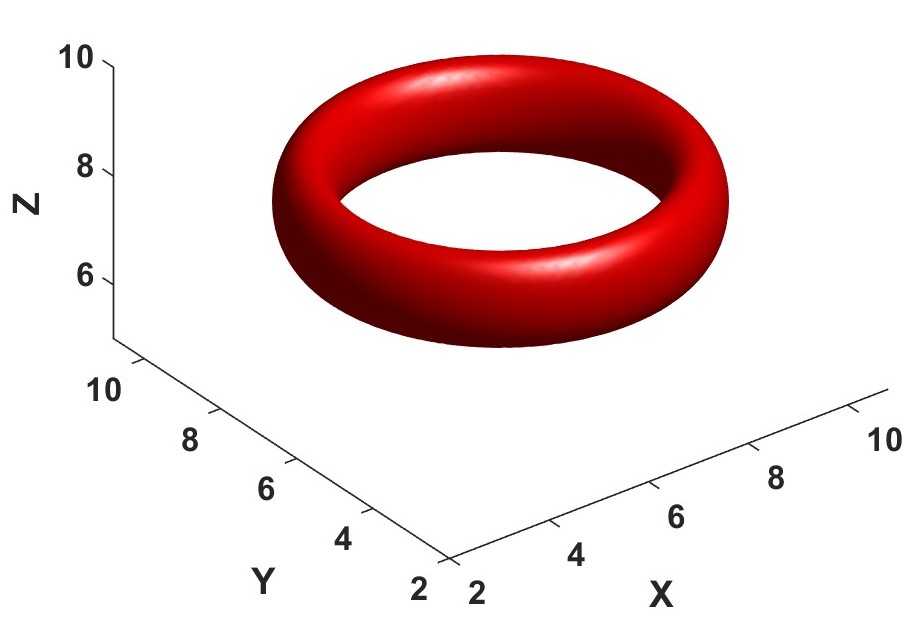} %
\xincludegraphics[width=0.45\linewidth, label={$\qquad$b)}]{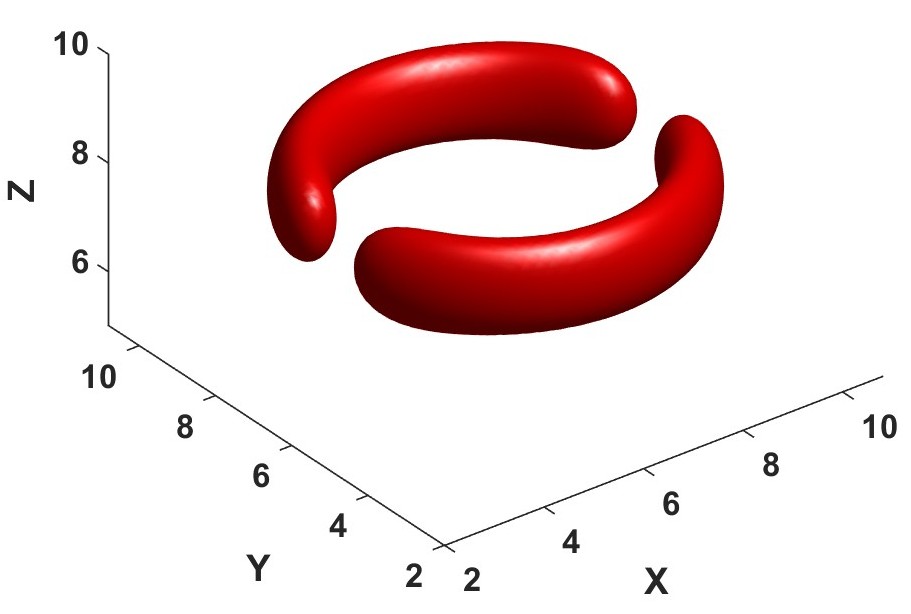} %
\xincludegraphics[width=0.45\linewidth, label={$\qquad$c)}]{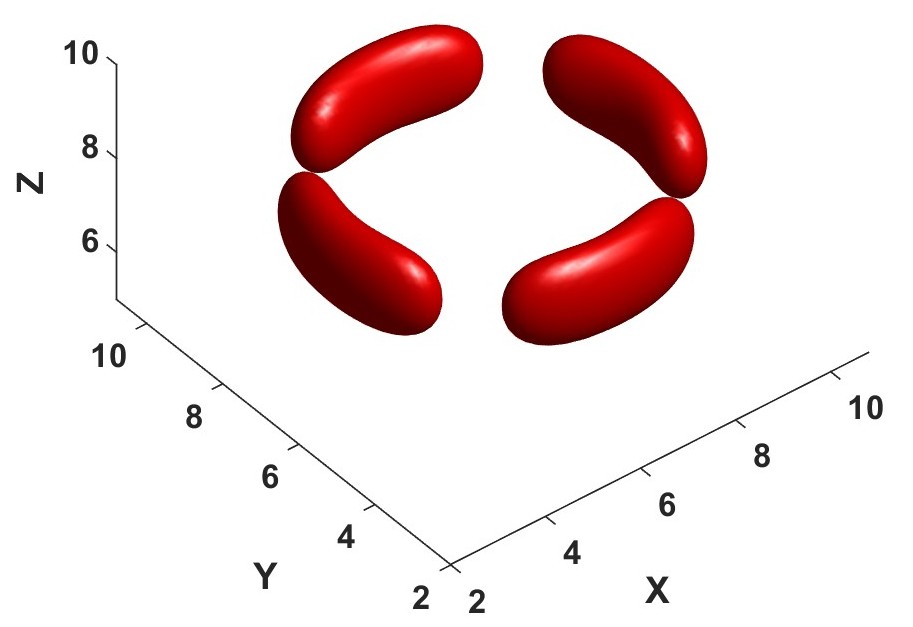} %
\xincludegraphics[width=0.45\linewidth, label={$\qquad$d)}]{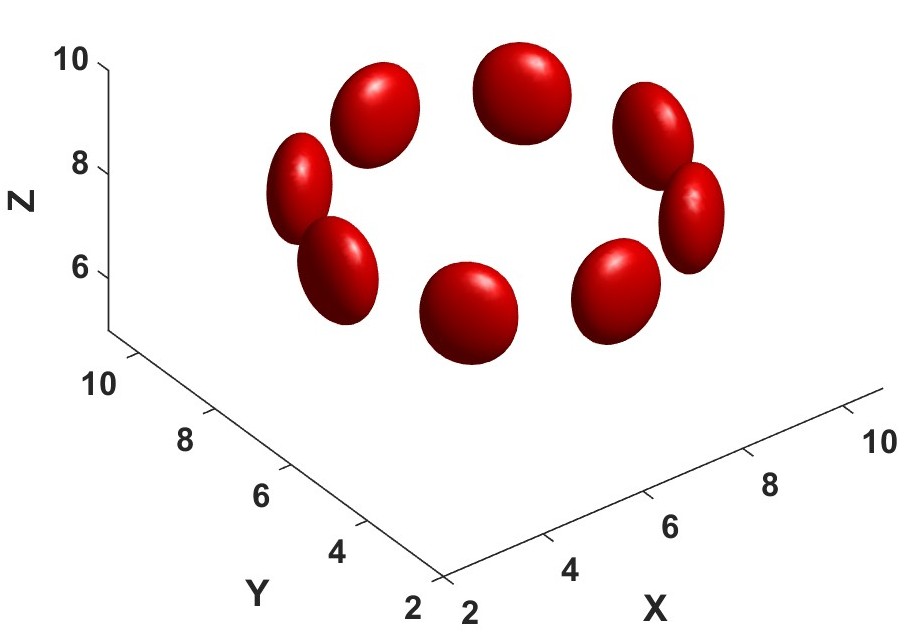} %
\xincludegraphics[width=0.45\linewidth, label={$\qquad$e)}]{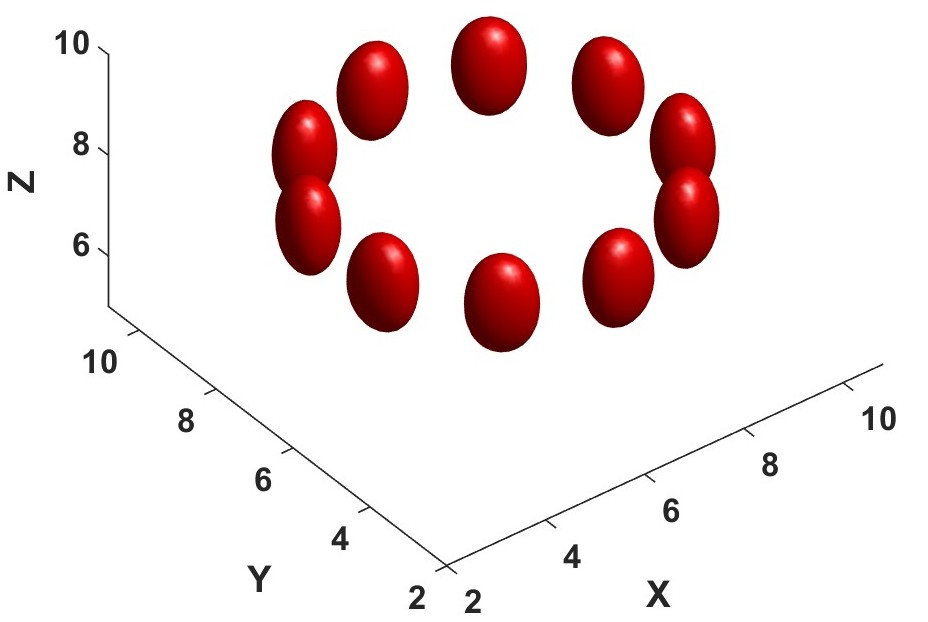} %
\xincludegraphics[width=0.45\linewidth, label={$\qquad$f)}]{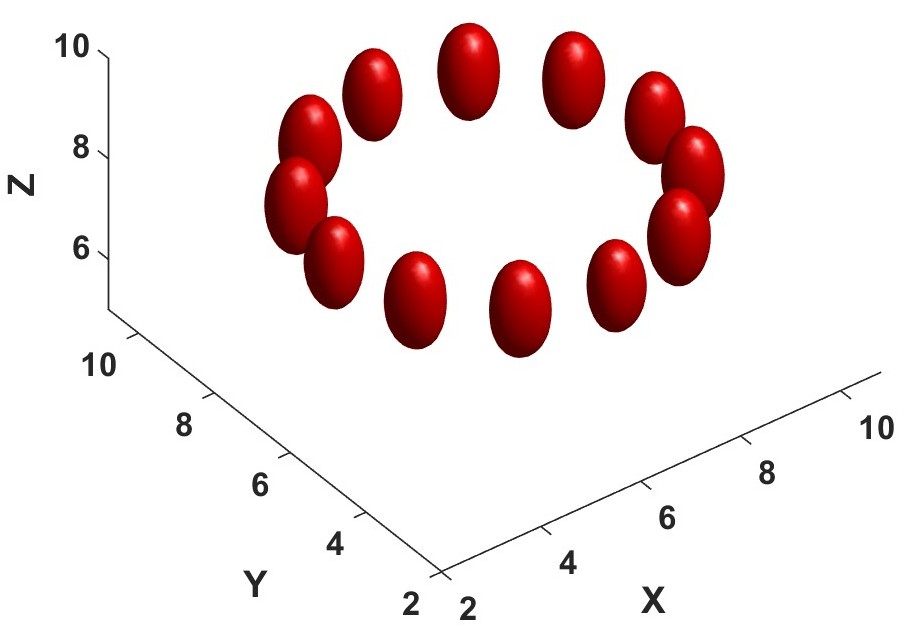}
\caption{Isosurface plots illustrating stable ring-shaped and multipole QD
density profiles for different winding numbers $S$, with norm $N=2000$,
DDI length $a_{\mathrm{dd}}=130.8a_{0}$, scattering length $a=100a_{0}$,
and the coefficient of the LHY term $\protect\gamma _{\mathrm{QF}%
}=2.697\times 10^{7}a_{0}^{5/2}$, confined in the toroidal potential (%
\protect\ref{toroid}). (a) The axisymmetric mode with $S=0$. (b) A dipole QD
configuration with $S=1$. (c) A quadrupole QD with four density lobes for $%
S=2$. The panels in the bottom row display stable higher-order multipole
structures with $8$, $10$, and $12$ density lobes, obtained for $S=4$, $S=5$%
, and $S=6$, respectively. All the stationary states are obtained by means
of the ITP method.}
\label{s_0}
\end{figure}

\section{The model}

\label{model} The dipolar BEC composed of $N$ atoms with mass $m$, whose
magnetic moment is oriented along the $z$-axis, is accurately modeled by the
LHY-amended GP equation, which includes the contact interactions and DDI
\cite{young2023dipole,adhikari2024quasi}:
\begin{eqnarray}
i\hbar \frac{\partial \psi (\mathbf{r},t)}{\partial t} &=&\bigg[-\frac{\hbar
^{2}}{2m}\nabla ^{2}+V(\mathbf{r})+\frac{4\pi \hbar ^{2}}{m}aN|\psi (\mathbf{%
r},t)|^{2}  \notag \\
&&+\frac{3\hbar ^{2}}{m}a_{\mathrm{dd}}N\int V_{\mathrm{dd}}(\mathbf{R}%
)|\psi (\mathbf{r}^{\prime})|^{2}d\mathbf{r}^{\prime }  \notag \\
&&+\frac{\gamma _{\mathrm{QF}}\hbar ^{2}N^{3/2}}{m}|\psi (\mathbf{r},t)|^{3}%
\bigg]\psi (\mathbf{r},t),  \label{gpe}
\end{eqnarray}%
\begin{equation}
V_{\mathrm{dd}}(\mathbf{R})=\frac{1-3\cos ^{2}\theta }{|\mathbf{R}|^{3}},~%
\mathbf{R}=\mathbf{r}-\mathbf{r}^{\prime },~a_{\mathrm{dd}}=\frac{\mu
_{0}\mu ^{2}m}{12\pi \hbar ^{2}},
\end{equation}%
where $\nabla ^{2}$ is the 3D Laplacian, $V(\mathbf{r})$ is the external
potential, $a$ denotes the scattering length of the contact interaction and $%
a_{\mathrm{dd}}$ denotes the dipolar interaction length. Kernel $V_{\mathrm{%
dd}}$ determines long-range interaction between two magnetic atoms
positioned at $\mathbf{r}=\{x,y,z\}$ and $\mathbf{r}^{\prime }=\{x^{\prime
},y^{\prime },z^{\prime }\}$, where $\theta $ is the angle between vector $%
\mathbf{R}$ and the $z\ $axis. The singularity introduced by the dipolar potential is taken care of by evaluating them in momentum space ($k$), where the momentum-space kernel was regularized using a spherical truncation method \cite{young2023openmp} and the $k=0$ component was set to zero, avoiding the singularity while preserving the long-range anisotropic character of the dipolar interaction. The wavefunction is normalized to $\int |\psi (\mathbf{r},t)|^{2}d\mathbf{r}=1$. In the scaled form, the effective DDI length is $\epsilon _{%
\mathrm{dd}}\equiv a_{\mathrm{dd}}/a$. For the formation of QDs, we consider
strongly dipolar atoms, with $a_{\mathrm{dd}}>a$ (for $a_{\mathrm{dd}}<a$,
the system is effectively repulsive, causing decay of localized inputs).
Taking the value of $a_{\mathrm{dd}}=130.8a_{0}$ \cite{ska} for $^{164}%
\mathrm{Dy}$ atoms, where $a_{0}$ is the Bohr radius, there remains
flexibility in determining the scattering length $a$, as it may be tuned by
means of the Feshbach resonance \cite{chin2010feshbach}. Here, we set $%
a=100a_{0}$. The coefficient of the LHY (alias quantum-fluctuation, QF) term
\cite{2025dynamically,young2023spontaneous,2025hollow} is
\begin{equation}
\gamma _{\mathrm{QF}}=\frac{128}{3}\sqrt{\pi a^{5}}Q_{l}(\epsilon _{\mathrm{%
dd}}),
\end{equation}%
with the auxiliary function \cite{young2024expansion}
\begin{equation}
Q_{l}(\epsilon _{\mathrm{dd}})=(1-\epsilon _{\mathrm{dd}})^{l/2}{_{2}}F_{1}%
\bigg(\frac{-l}{2},\frac{1}{2};\frac{3}{2};\frac{3\epsilon _{\mathrm{dd}}}{%
\epsilon _{\mathrm{dd}}-1}\bigg)
\label{hypergeo}
\end{equation}%
where ${_{2}}F_{1}\bigg(\dfrac{-l}{2},\dfrac{1}{2};\dfrac{3}{2};\dfrac{%
3\epsilon _{\mathrm{dd}}}{\epsilon _{\mathrm{dd}}-1}\bigg)$ is the
hypergeometric function. Here, $l$ is an integer, represents the order of the momentum-space integral in the Bogoliubov treatment of QF. Following \cite{lima2011quantum}, the effect of the DDI on the QF-induced energy correction is most significant for $l=5$, which is therefore used in the evaluation of $Q_l(\epsilon_{\mathrm{dd}})$ in our calculations, improving the prospects for experimental verification. In this
case, Eq. (\ref{hypergeo}) reduces to
\begin{equation}
Q_{5}(\epsilon _{dd})\approx 1+\frac{3}{2}\epsilon _{\mathrm{dd}}^{2}
\label{q5}
\end{equation}%
\cite{bisset2016ground}. We use the simplified expression of $Q_{5}(\epsilon
_{\mathrm{dd}})$ (Eq. (\ref{q5})) in numerical simulations, for which $%
\gamma _{\mathrm{QF}}=2.697\times 10^{7}a_{0}^{5/2}$. The scaled form of Eq.
(\ref{gpe}) is produced by measuring lengths in units of $l_{0}=\sqrt{\hbar
/m\omega _{0}}$, time and energy in units of $t_{0}=\omega _{0}^{-1}$ and $%
\hbar \omega _{0}$, respectively, and density $|\psi |^{2}$ in units of $%
l^{-3}$ (here $\omega _{0}$ is the reference angular frequency, introduced
for the nondimensionalization of the GP equation; it is not a frequency
imposed by the toroidal trapping potential in Eq.~(\ref{toroid})) :
\begin{eqnarray}
i\frac{\partial \psi (\mathbf{r},t)}{\partial t} &=&\bigg[-\frac{1}{2}\nabla
^{2}+V(\mathbf{r})+4\pi aN|\psi (\mathbf{r},t)|^{2}  \notag \\
&&+3a_{\mathrm{dd}}N\int \frac{1-3\cos ^{2}\theta }{|\mathbf{R}|^{3}}|\psi (%
\mathbf{r}^{\prime })|^2d\mathbf{r}^{\prime }  \notag \\
&&+\gamma _{\mathrm{QF}}N^{3/2}|\psi (\mathbf{r},t)|^{3}\bigg]\psi (\mathbf{r
},t).  \label{dgpe}
\end{eqnarray}
\begin{figure}[tbp]
\centering
\xincludegraphics[width=0.45\linewidth, label={$\quad$a)}]{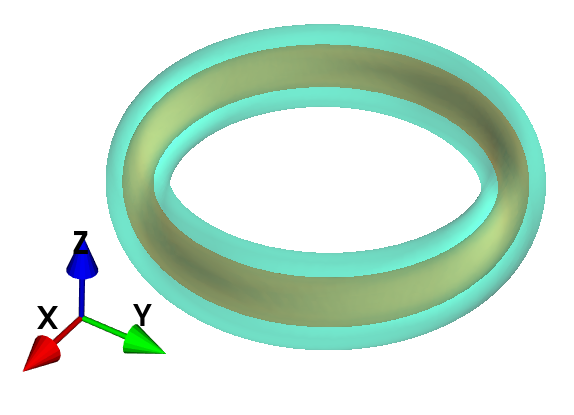} %
\xincludegraphics[width=0.45\linewidth, label={$\quad$b)}]{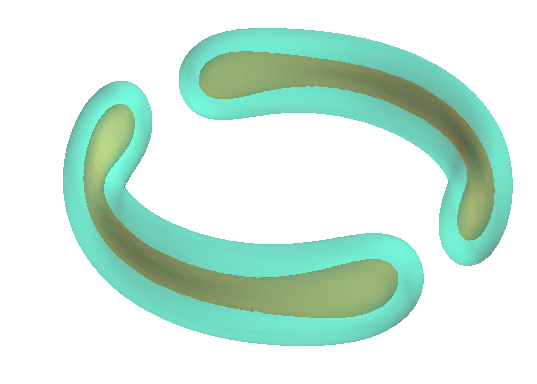} %
\xincludegraphics[width=0.45\linewidth, label={$\quad$c)}]{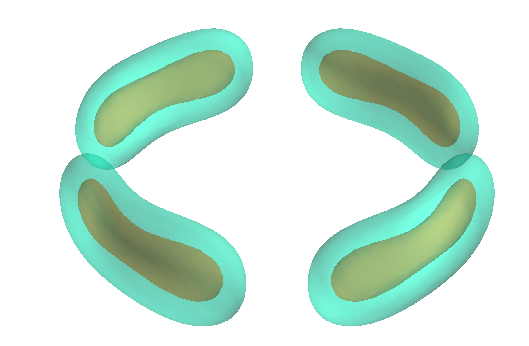} %
\xincludegraphics[width=0.45\linewidth, label={$\quad$d)}]{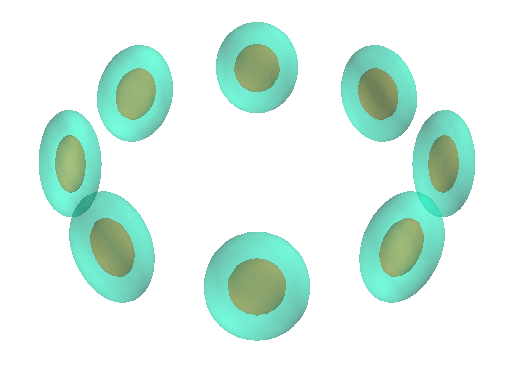} %
\xincludegraphics[width=0.45\linewidth, label={$\quad$e)}]{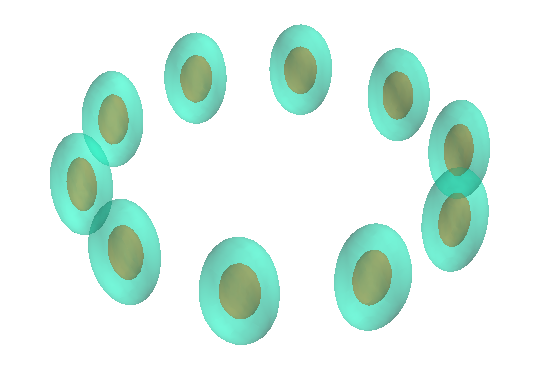} %
\xincludegraphics[width=0.45\linewidth, label={$\quad$f)}]{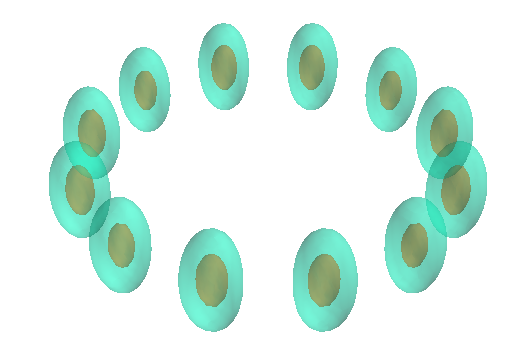}
\caption{The magnified view of the density distribution in the stable
stationary ring-shaped and multipole QDs in the absence of the LHY
correction, $\protect\gamma _{\mathrm{QF}}=0$, for different vorticities.
Results are shown for $N=5000$ atoms with DDI length $a_{\mathrm{dd}%
}=130.8a_{0}$ and scattering length $a=100a_{0}$, confined in the toroidal
potential (\protect\ref{toroid}). Panels correspond to: (a) $S=0$, (b) $S=1$,
(c) $S=2$, (d) $S=4$, (e) $S=5$, and (f) $S=6$. The plots display the spatial
localization and symmetry of the stable QDs. The stationary states are
obtained by means of the ITP method.}
\label{inner_density}
\end{figure}
We here adopt the natural form of the toroidal trapping potential,
which is available in the experiment \cite{torus1,torus2}:
\begin{gather}
V(\mathbf{r})=-p\exp \bigg[-\frac{(\rho-\rho_{0})^{2}}{d^{2}}-\frac{z^{2}}{z_{0}^{2}}\bigg],  \label{toroid} \\
\rho \equiv \sqrt{x^{2}+y^{2}},  \notag
\end{gather}
where coefficients $p$ and $\rho_{0}$, $d$, $z_{0}$ define the
potential depth, and obvious geometric parameters,
respectively. Here, $p$ is expressed in terms of $\hbar \omega
_{0}$.
The mass of $^{164}$Dy atoms is $m=164\times 1.66054\times 10^{-27}$
kg. With the reference frequency $\omega _{0}=2\pi \times 61$ Hz, the
characteristic length and time are $l_{0}=\sqrt{\hbar /m\omega _{0}}\equiv
1.01~\mu $m and $t_{0}=\omega _{0}^{-1}\equiv 2.61$ ms. Further, $\rho _{0}$, $d
$, $z_{0}$ are expressed in terms of the above scaling parameter $l_{0}$.
Expanding Eq. (\ref{toroid}) around its minimum ($\rho=\rho_{0}$, $z=0$), yields
the following values of the second derivatives
\begin{eqnarray}
V''(\rho_{0})=\frac{2p}{d^{2}}, \qquad  V''(0)=\frac{2p}{z_{0}^{2}}.
\end{eqnarray}
For the trap parameters $p=4$, $\rho_{0}=2\pi $, $d=\sqrt{6}$, and $%
z_{0}=\sqrt{6}$, we thus obtain
$V''(\rho_{0})=V''(0)=4/3$, hence the effective radial and axial trapping
frequencies (in units of $\omega_{0}$) are
\[
\Omega_{\rho}=\Omega_{z}=\Omega=\sqrt{4/3}.
\]
Using the reference frequency $\omega_{0}=2\pi\times 61$ Hz, the respective
physical-trap frequencies are $f_{\rho}=f_{z}=\Omega\times (\omega_{0}/2\pi)
\simeq 70.43~\mathrm{Hz}$. Unless stated otherwise, the numerical results
are presented using the above-mentioned
trapping parameters, which adequately represent the generic case (simulations
performed with other parameters produced similar results).
The study of multipole QDs is carried out by varying atom number $N$ and the
vorticity $S$ (the topological charge, alias the winding number), while
fixing the DDI length $a_{\mathrm{dd}}=130.8a_{0}$, the scattering
length of the contact interaction $a=100a_{0}$, and the LHY coefficient $%
\gamma _{\mathrm{QF}}=2.697\times 10^{7}a_{0}^{5/2}$. The scaled equation (%
\ref{dgpe}) keeps the unitary normalization, i.e.,
\begin{equation}
\int \int \int |\psi |^{2}dxdydz=1.
\end{equation}

Using the variational principle, Eq. (\ref{gpe}) can be written as
\begin{equation}
i\frac{\partial \psi }{\partial t}=\frac{\delta E}{\delta \psi ^{\ast }},
\end{equation}%
where $\delta /\delta \psi ^{\ast }$ is the variational derivative, and the
system's energy is
\begin{eqnarray}
E &=&\int d\mathbf{r}\bigg[\frac{1}{2}|\nabla \psi (\mathbf{r})|^{2}+V(r)|\psi (\mathbf{%
r})|^{2}+2\pi Na|\psi (\mathbf{r})|^{4}  \notag \\
&&+\frac{3}{2}a_{\mathrm{dd}}N|\psi (\mathbf{r})|^{2}\int \frac{1-3\cos
^{2}\theta }{|\mathbf{R}|^{3}}|\psi (\mathbf{r}^{\prime })|^{2}d\mathbf{r}%
^{\prime }  \notag \\
&&+\frac{2\gamma _{\mathrm{QF}}}{5}N^{3/2}|\psi (\mathbf{r})|^{5}\bigg].
\end{eqnarray}%

\section{Numerical results}

\label{results} The integrodifferential GP equation Eq. (\ref{dgpe}) was
solved numerically using the split-time-step Crank-Nicolson algorithm,
adapting the FORTRAN/C programs \cite{kumar2015fortran} or their
open-multiprocessing versions \cite{lonvcar2016openmp}. In particular,
stationary solutions were obtained by means of the ITP method. This was
performed with nearly equal discretization steps in the $x$, $y$, and $z$
directions.

Droplets confined in the toroidal potential resemble lifebelts (or \textquotedblleft donuts").
We address VQDs solutions with the vorticity up to $S=12$. The trapping
toroidal potential significantly constrains the VQDs, unlike the
characteristic flat-top profiles of free-space VQDs and vortex solitons \cite%
{li2018two,berezhiani2001dynamics}. To enable the numerical investigation of
the VQDs states, we use the input shaped as the phase-imprinted Gaussian
states \cite{young2025giant}:
\begin{equation}
\psi (x,y,z)\propto\left( x+i\,y\right) ^{S}\exp \left( -\frac{\rho^{2}}{\delta
_{r}^{2}}-\frac{z^{2}}{\delta _{z}^{2}}\right) ,\qquad  \label{input}
\end{equation}%
where $\rho^{2}=x^{2}+y^{2}$. The phase of this complex expression is $%
\Phi=S\arctan (y,x)$. $\delta _{r}$ and $\delta _{z}$ are the width parameters,
we set $\delta _{r}=\delta _{z}\equiv \sqrt{2}$ for the numerical
calculation.
\begin{figure}
\centering
\includegraphics[width=0.8\linewidth]{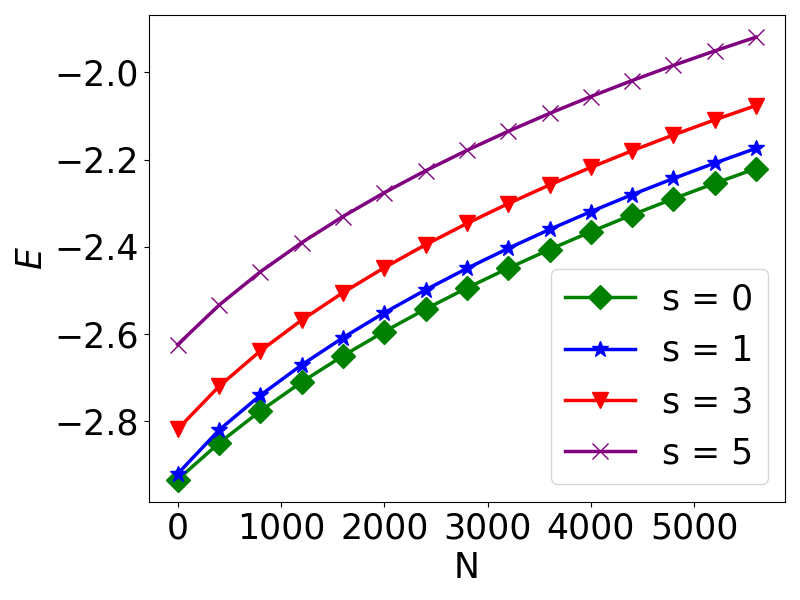}
\caption{Total energy $E$ as a function of atom number $N$ for different
vorticities $S$ in the absence of the LHY correction
($\gamma_{\mathrm{QF}}=0$), the dipolar and scattering lengths being
$a_{\mathrm{dd}} = 130.8\,a_{0}$ and $a = 100\,a_{0}$. The stationary states are
produced by means of ITP. With the increase of $N$, the energy increases monotonously for
given values of $S$. The $S=0$ configuration yields the lowest energy,
corresponding to the ground state, while the branches with $S=1,3,5$
correspond to multipole excited states.}
\label{E_vs_N_lhyfree}
\end{figure}
\begin{figure}[tbp]
\centering
\xincludegraphics[width=0.49\linewidth,label={a)}]{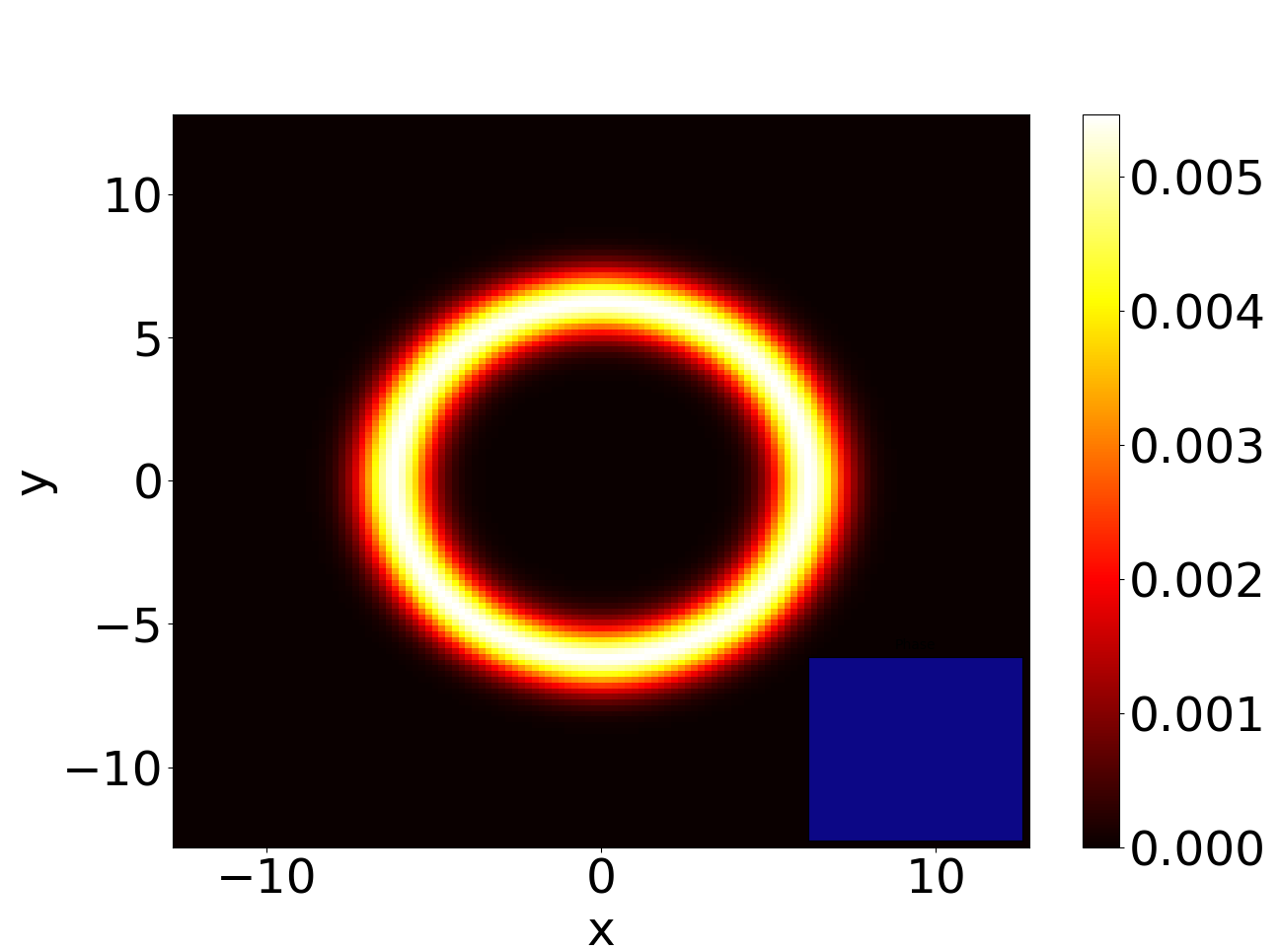} %
\xincludegraphics[width=0.49\linewidth, label={b)}]{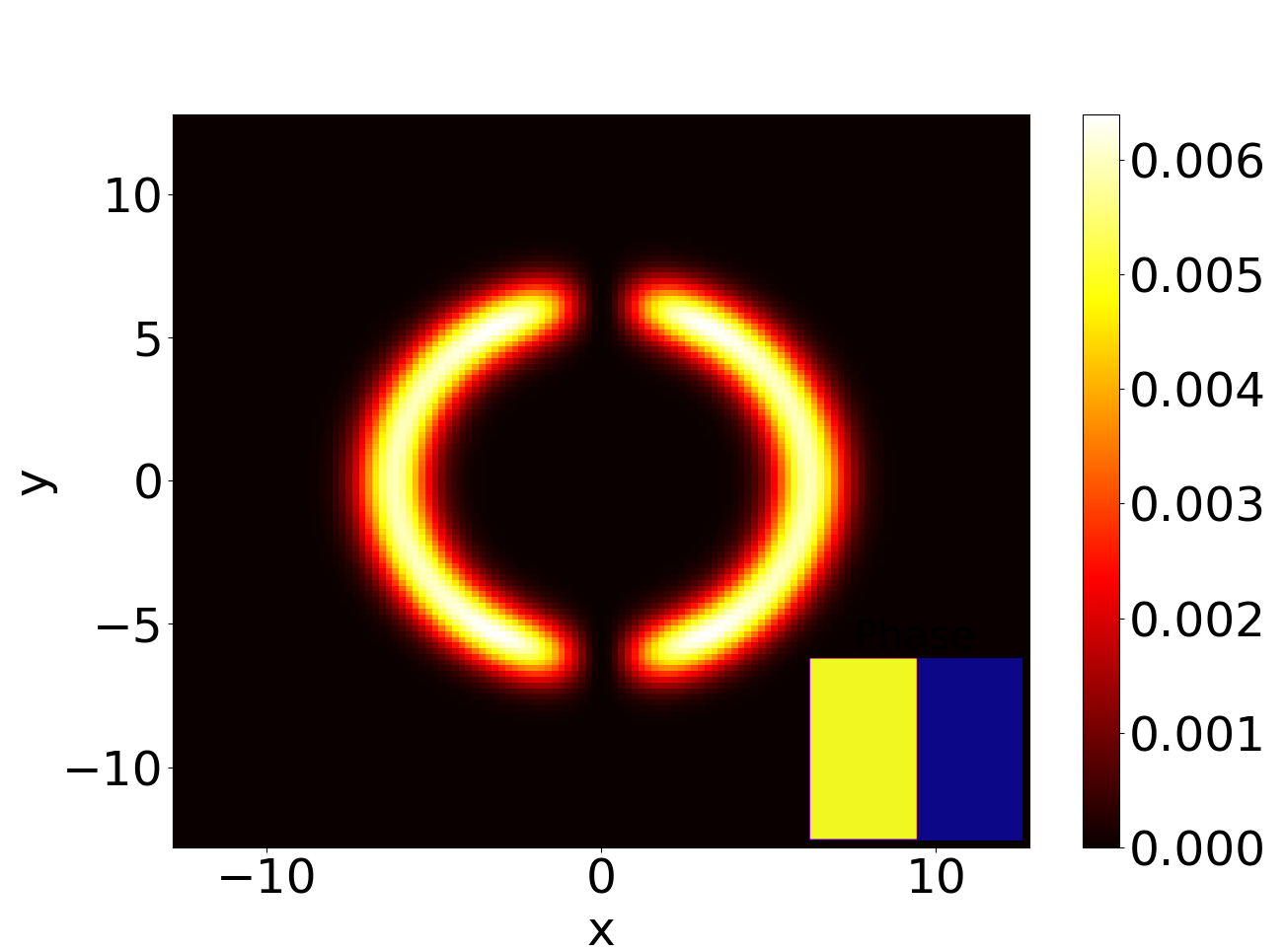}
\xincludegraphics[width=0.49\linewidth, label={c)}]{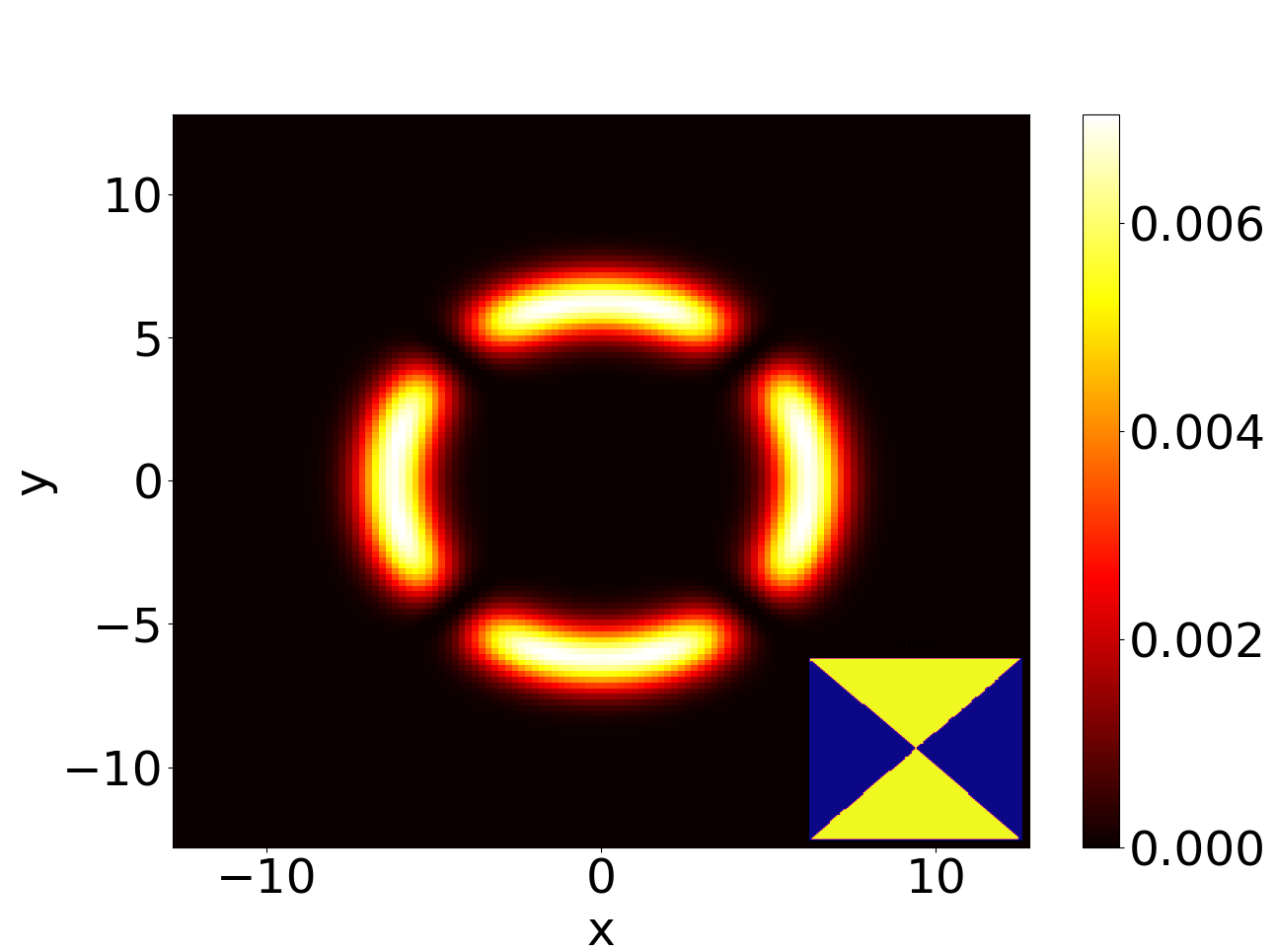} %
\xincludegraphics[width=0.49\linewidth, label={d)}]{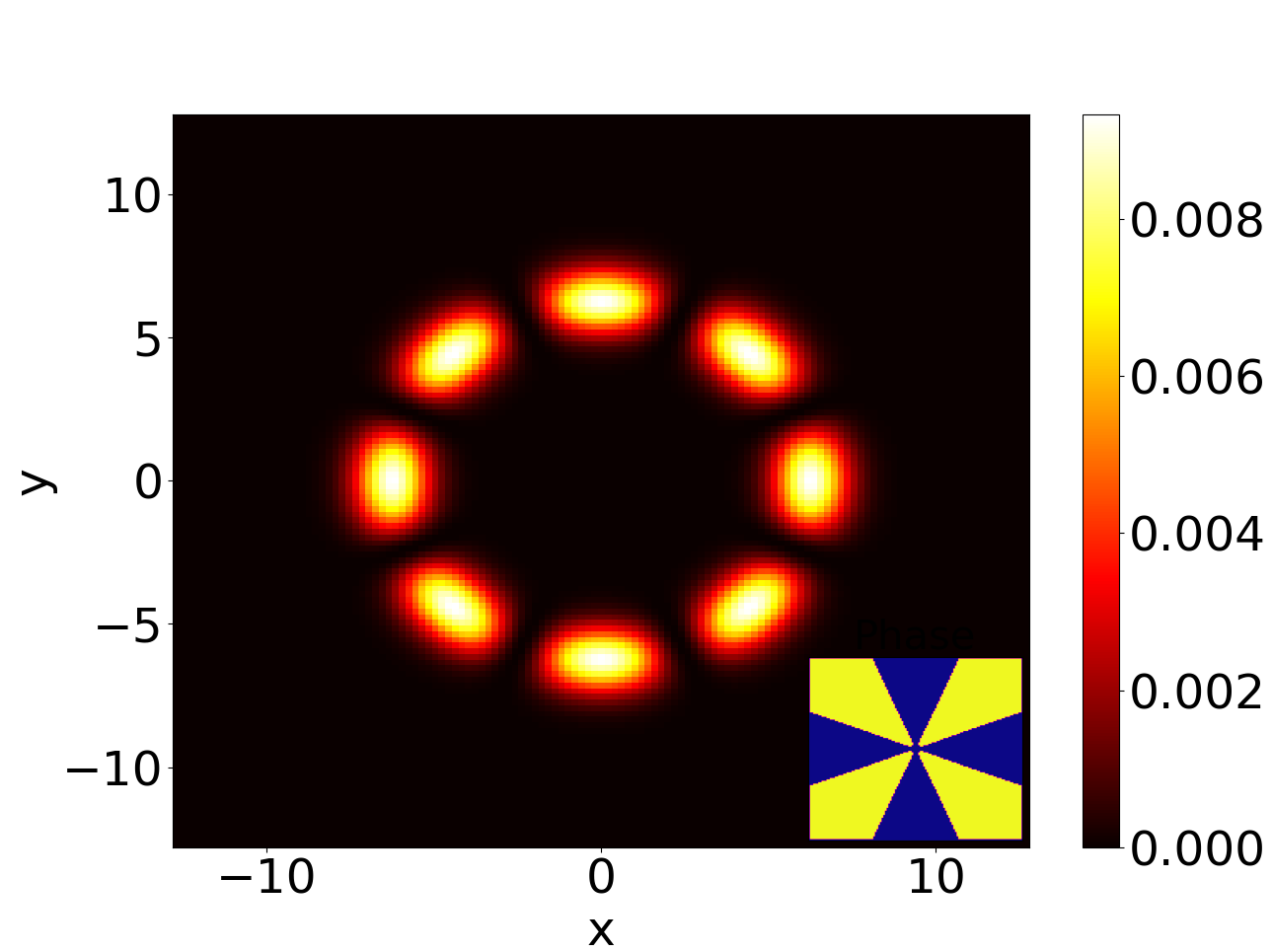}
\xincludegraphics[width=0.49\linewidth, label={e)}]{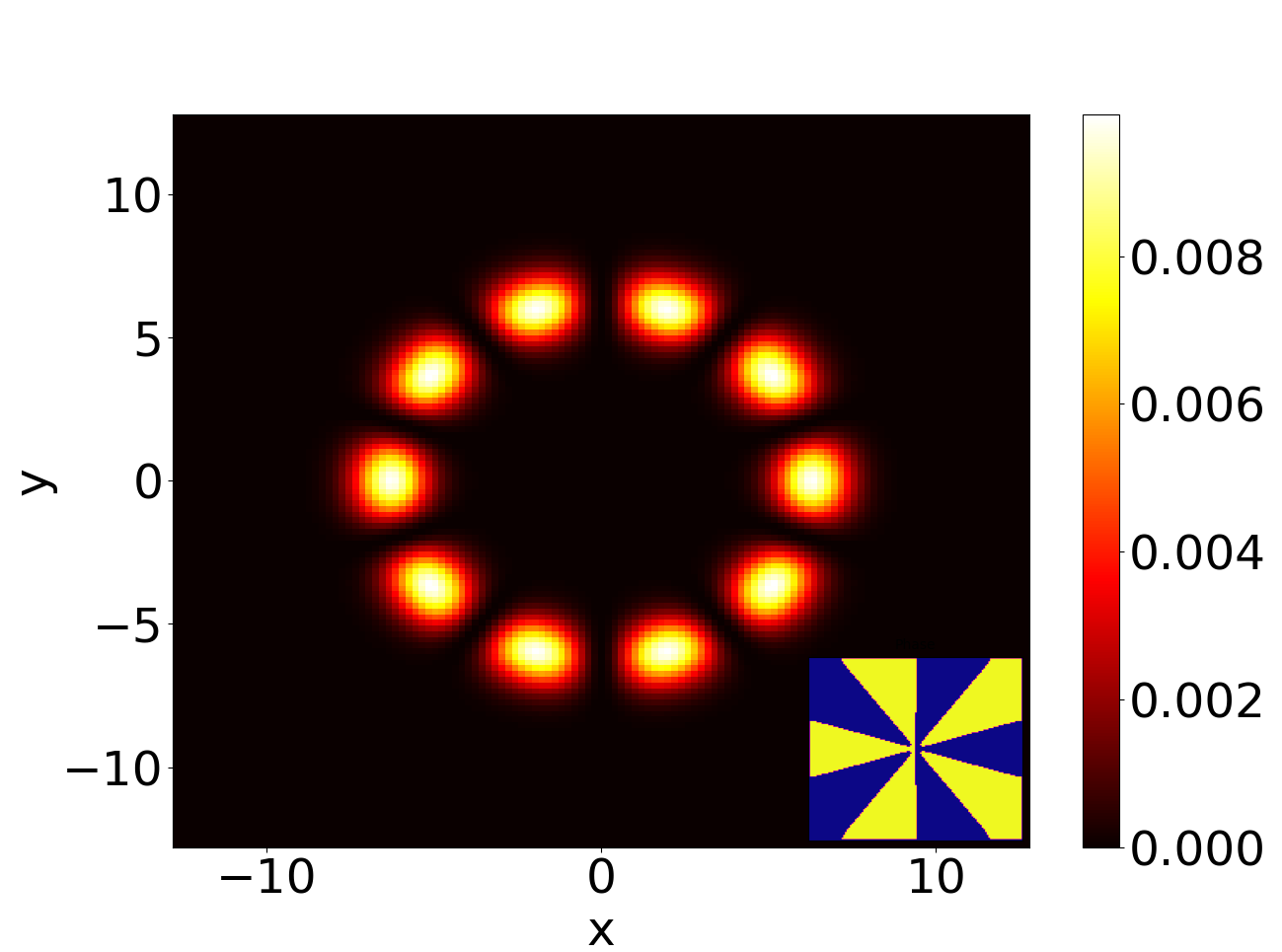} %
\xincludegraphics[width=0.49\linewidth, label={f)}]{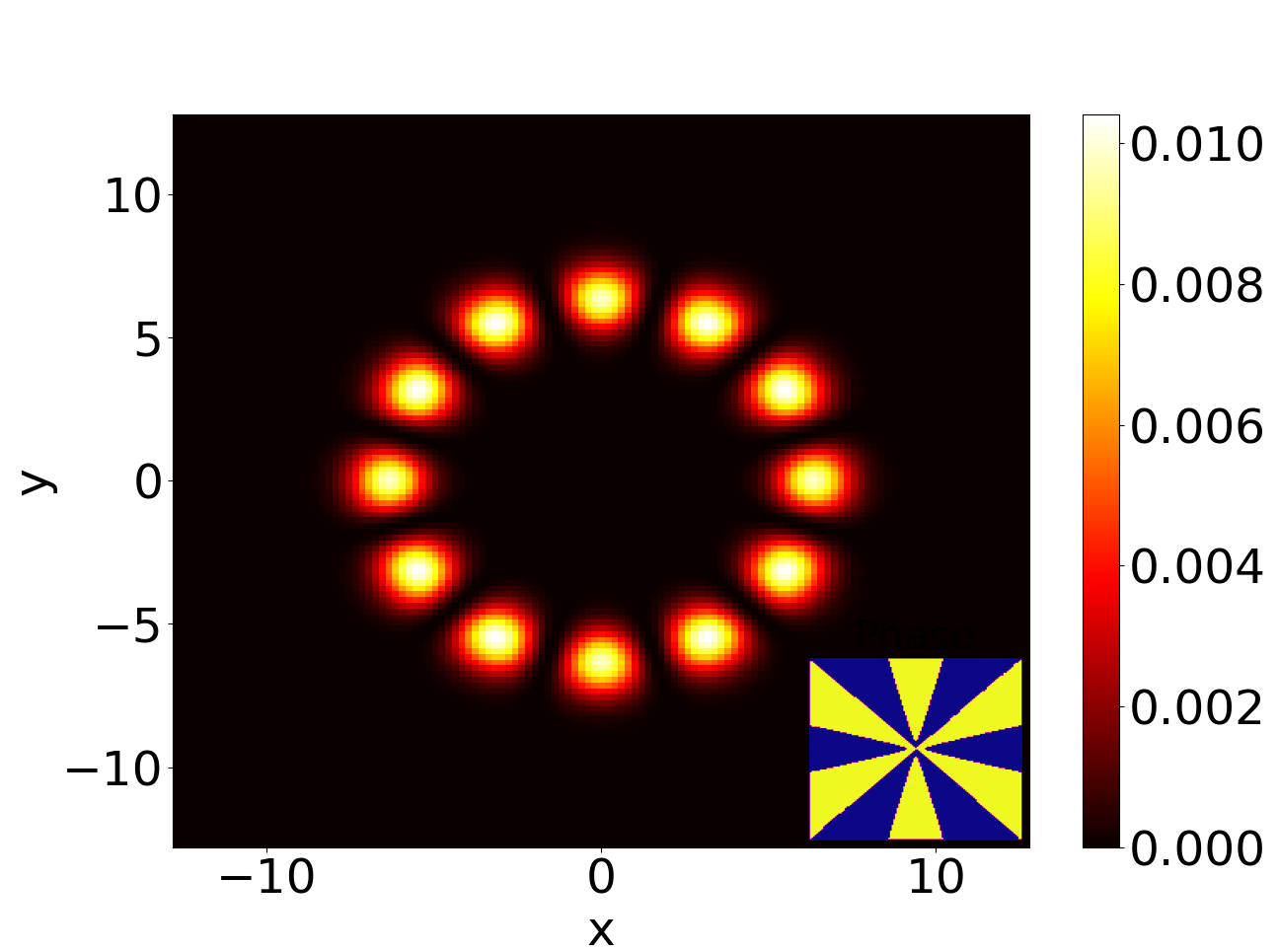}
\caption{Contour plots of the 2D density projection in the $(x,y)$ plane,
corresponding to the stable stationary isosurface profiles, shown in Fig.~%
\protect\ref{s_0}, for $N=2000$, $a_{dd}=130.8a_{0}$, $a=100a_{0}$, $%
\protect\gamma _{QF}=2.697\times 10^{7}a_{0}^{5/2}$ and for different vorticities
$S$: (a) $S=0$, (b) $S=1$, (c) $S=2$, (d) $S=4$, (e) $S=5$, and (f) $S=6$.
The inset in each panel shows the corresponding phase profile, which
reveals the topological structure and phase singularities of each state. All
stationary configurations were produced by means of the ITP method.}
\label{projection}
\end{figure}

\subsection{Stationary states of multipole droplets under toroidal
confinement}

\label{stat_ste}
\begin{figure*}[tbp]
\centering
\xincludegraphics[width=0.31\linewidth, label={a)},pos={n}]{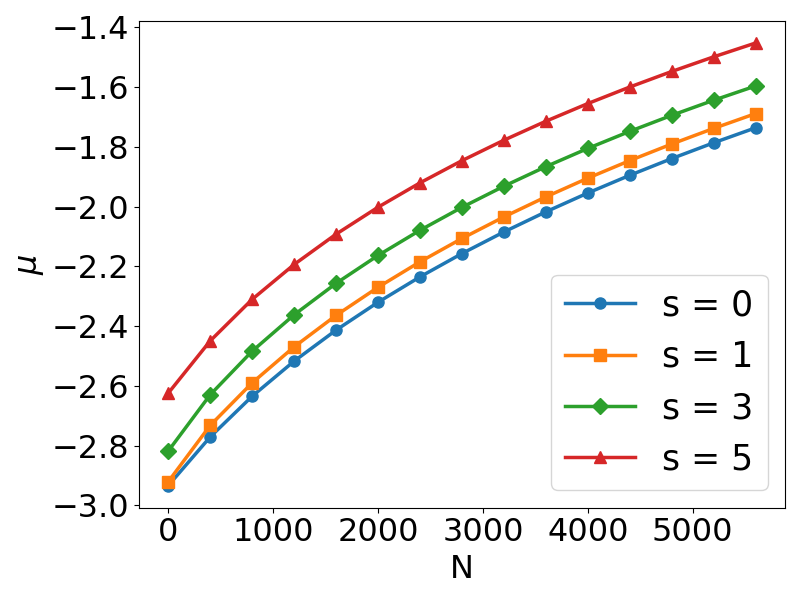} %
\xincludegraphics[width=0.31\linewidth, label={b)},pos={n}]{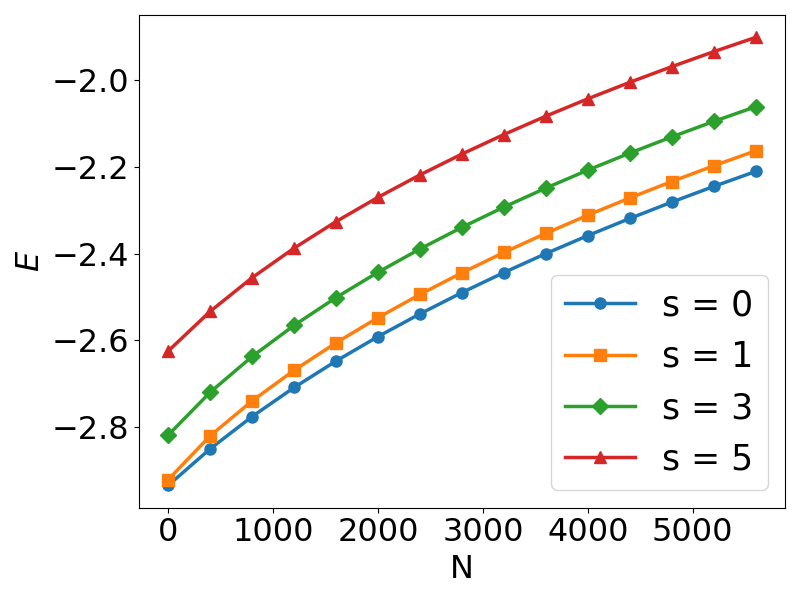} %
\xincludegraphics[width=0.31\linewidth, label={c)},pos={n}]{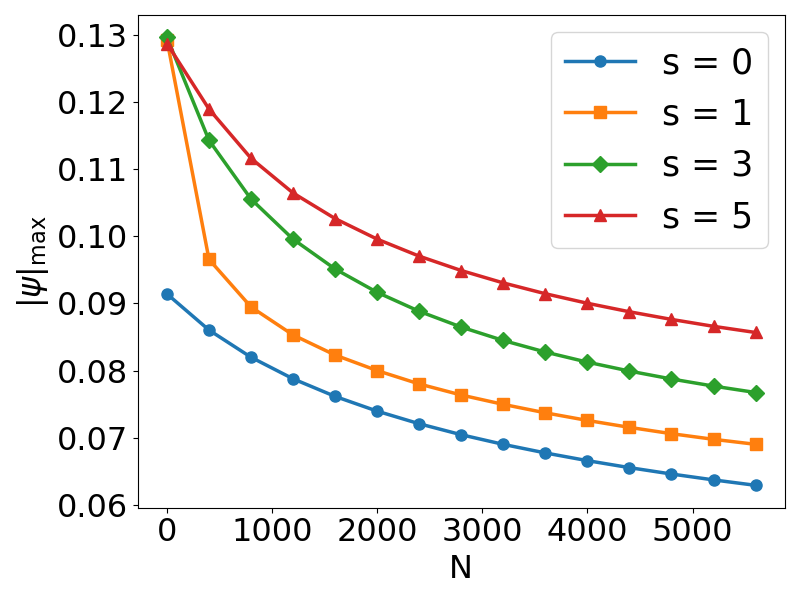}
\caption{a) The chemical potential $\protect\mu $, (b) the total energy $E$,
and (c) amplitude $\protect\psi _{\max }$ vs. the number of atoms $N$ in
stationary QD\ states with DDI length $a_{\mathrm{dd}}=130.8a_{0}$,
scattering length $a=100a_{0}$, and LHY coefficient $\protect\gamma _{%
\mathrm{QF}}=2.697\times 10^{7}a_{0}^{5/2}$, for vorticities $S=0,1,3,5$
under the action of the toroidal confining potential (\protect\ref{toroid}).
All the configurations presented here are stationary solutions obtained by
means of the ITP method. As $N$ increases, both $\protect\mu $ and energy
increase monotonously for all multipole states, reflecting the increase of
the repulsive-interaction energy. In contrast, the amplitude decreases due
to the spreading of the respective wavefunctions. States with higher values
of $S$ exhibit larger energy and chemical potential for given $N$, and their
amplitudes are consistently higher.}
\label{natoms}
\end{figure*}
The interplay of the toroidal confinement, represented by
Eq. (\ref{toroid}), with the local and nonlocal interactions included in Eq.
(\ref{dgpe}) leads to diverse morphologies of the VQDs. Naturally, multipole
droplets tend to form with an even number of lobes, due to the phase-winding
constraints. The origin of these multipole structures can be traced to the centrifugal
force. For a vortex of charge $S$, the wave function
$\psi=f(\rho,z)e^{iS\theta}$ generates a phase-gradient kinetic-energy term
$S^{2}/(2\rho^{2})$, which acts as an effective centrifugal potential,
independent of density gradients, hence it is present even in the
Thomas–Fermi regime. As $S$ increases, the centrifugal pressure pushes the
density outward and destabilizes the uniform annular profile. Typical stationary density profiles for different vorticity
values are presented in Fig.~\ref{s_0}. For $S=0$, a ring-shaped structure
appears, as shown in Fig.~\ref{s_0}(a). When vorticity $S=1$ is imparted, the
centrifugal force causes the droplet to break the rotational symmetry and
split in two distinct lobes, forming a dipole-like configuration, see Fig.~%
\ref{s_0}(b). As the vorticity increases to $S=2$, the VQD give rise to
quadrupole-shaped modes, see Fig.~\ref{s_0}(c). The growing centrifugal
effect pushes the density outwards along the radial direction. By further
increasing the vorticity to $S=4$, $5$ and $6$, higher-order multipole
structures emerge, consisting of $8$, $10$ and $12$ density lobes,
respectively, as displayed in panels (d), (e), and (f) of Fig. \ref{s_0}. The
lobes arrange themselves symmetrically around the radial axis, forming a
necklace-like distribution as a result of the interplay of the phase
winding, centrifugal expansion, and toroidal confinement. Figure~\ref{inner_density}
presents a magnified view of the density
distribution in QDs composed of $N=5000$ atoms in the absence of the
LHY correction, $\gamma_{\mathrm{QF}}=0$. Although the obtained stationary
profiles are similar to the LHY-stabilized states shown in
Fig.~\ref{s_0}, they are not self-bound droplets; instead,
they constitute trap-bound stationary solutions maintained
by the combined effect of the DDI anisotropy \cite{zajec2015mean} and toroidal trapping potential.
To clarify their energy character, the total energy is shown in Fig.~\ref{E_vs_N_lhyfree} as a function of $N$
for fixed vorticities $S$.
The energy curves show that the ring-shaped state with $S=0$ attains the lowest energy
and therefore represents the ground state,
while multipole configurations ($S=1,3,5$) correspond to local extrema (metastable states).
Thus, the LHY-free stationary profiles are regarded as ITP-generated
solutions supported by the DDI anisotropy and the toroidal geometry, rather than
as globally stable or long-lived droplets. We further emphasize that the
energy metastability does not guarantee full dynamical stability. It is well known that
dipolar droplets are not dynamically stable in the absence of the LHY term. The full stability
analysis is therefore presented here only for the LHY-stabilized states, which are
the main subject of the present work. Both Figs. \ref{s_0} and \ref%
{inner_density} demonstrates that the number $n$ of the density lobes (poles
of the multipole patterns) increases with the increase of vorticity $S$. For
$S\leq 6$, the trend approximately follows an empirical relation, $n=2S$.
However, this correspondence is not a universal rule and tends to deviate
for higher vorticities, where the interplay between the centrifugal
expansion, nonlinear interactions, and confinement geometry leads to more
complex droplet distributions.
\begin{figure*}[tbp]
\centering
\xincludegraphics[width=0.4\linewidth, label={a)},pos={n}]{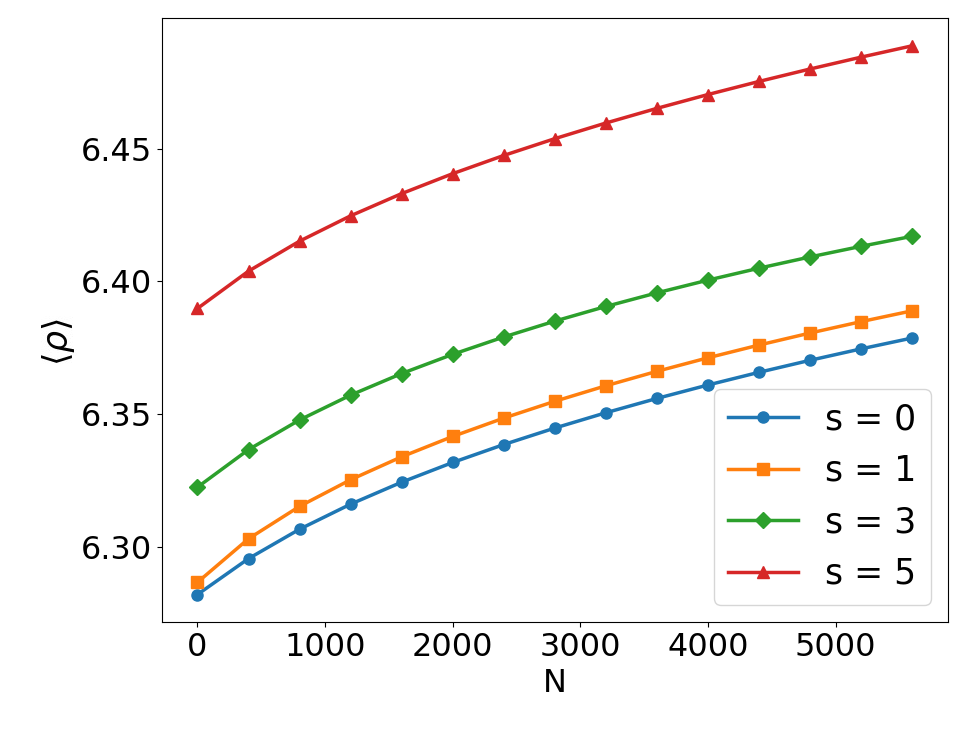} %
\xincludegraphics[width=0.4\linewidth, label={b)},pos={n}]{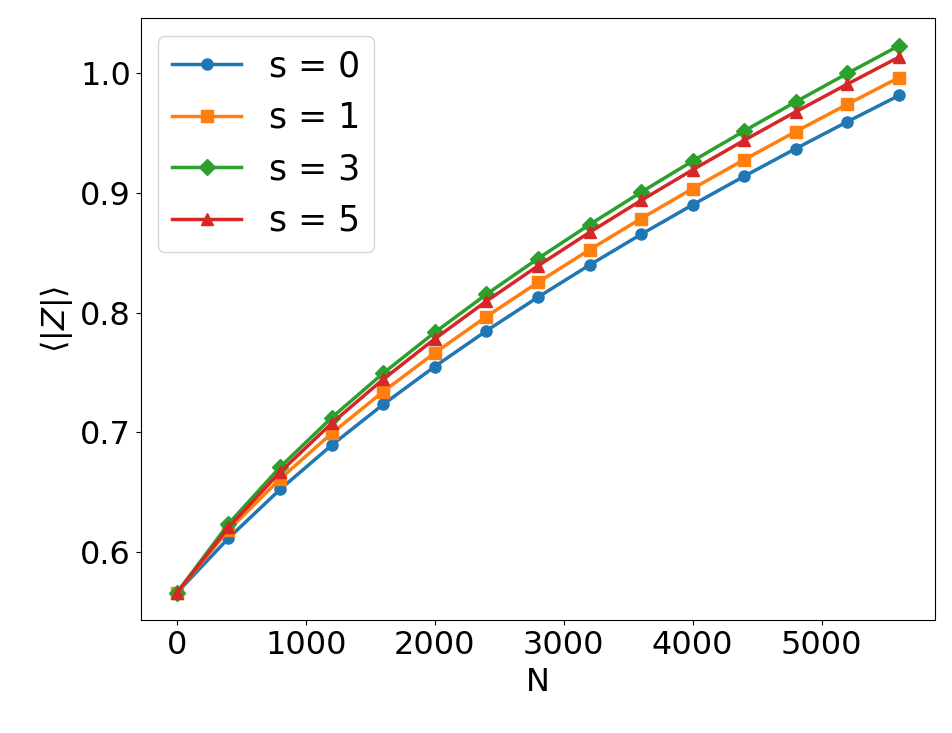}
\caption{(a) Radial extension $\langle \protect\rho \rangle $ and (b) axial
thickness $\langle |z|\rangle $ as functions of atom number $N$, for
different vorticities $S=0,1,3,5$, as calculated for the DDI length $a_{%
\mathrm{dd}}=130.8a_{0}$, scattering length $a=100a_{0}$, and $\protect%
\gamma _{\mathrm{QF}}=2.697\times 10^{7}a_{0}^{5/2}$. All the configurations
are stationary solutions obtained by means of the ITP method. The radial
size increases significantly with the growth of $N$ and $S$ due to the
centrifugal expansion of the rotating states. In contrast, the axial
thickness grows slower and shows less sensitivity to the vorticity,
reflecting the stronger confinement in the axial direction.}
\label{thickness}
\end{figure*}
Figure~\ref{projection} represents contour plots of the 2D density profiles
in the $\left( x,y\right) $ plane for the multipole QDs of the same types as
displayed in Fig.~\ref{s_0}. The panels correspond to: (a) $S=0$, (b) $S=1$,
(c) $S=2$, (d) $S=4$, (e) $S=5$, and (f) $S=6$, with the atom number fixed
at $N=2000$. The insets display the corresponding phase profiles for each
configuration, highlighting the phase winding and the topological structure
of the droplets. The phase profiles of these multipole lobes exhibit
out-of-phase behavior along the $z$-direction. The lobes feature alternating
phase patterns, which results in effective repulsive interactions. It should
be noted that the configurations presented in Figs.~(\ref{s_0}-\ref%
{projection}) are stationary solutions obtained by means of the ITP method.
Their existence does not automatically imply dynamical stability, which is
tested below using the perturbed real-time propagation. Proceeding to
characteristics of families of the QD modes, Fig.~\ref{natoms}(a) presents
the chemical potential $\mu $ as a function of the atom number $N$ for
distinct vorticity values. For all the vorticities considered, $\mu $
exhibits a positively slope the $\mu (N)$ dependence, which is consistent
with the anti-Vakhitov-Kolokolov (anti-VK) stability criterion, $d\mu /dN>0$
\cite{HS}. The usual VK criterion, $d\mu /dN<0$, is the necessary stability
condition in systems dominated by the self-attractive nonlinearity \cite%
{VK,Berge}. It is seen that the inclusion of the LHY correction universally
stabilizes the system, while increasing its chemical potential and energy.
Further, Fig. ~\ref{natoms}(b) displays the total energy as a function of $N$%
. As expected for the condensate with the repulsive interaction, the energy
increases with the growth of the atom number. The increase rate is higher
for multipolar QDs states due to their additional gradient energy.

Figure~\ref{natoms}(c) shows the amplitude as a function of the atom number $%
N$ for various vorticities. In contrast to the trends observed for the
chemical potential and energy, the amplitude decreases with the increase of $%
N$ for all vorticities. This behavior results from the enhanced centrifugal
effect associated with higher vorticity, which causes the condensate to
expand radially, thereby lowering the amplitude.
To understand the 3D structure of QDs under the action of the toroidal
confinement, we computed the average radial extension $\langle \rho \rangle$
and axial thickness $\langle |z|\rangle $, as per Eq. (\ref{rad_and_axi}),
as functions of the total atom number $N$ for distinct vorticity values $S$,
\begin{subequations}
\label{rad_and_axi}
\begin{align}
\langle \rho \rangle & =\int \!\!\!\int \!\!\!\int \sqrt{x^{2}+y^{2}}\,|\psi
(x,y,z)|^{2}\,dx\,dy\,dz,  \label{radial_width} \\
\langle |z|\rangle & =\int \!\!\!\int \!\!\!\int |z|\,|\psi
(x,y,z)|^{2}\,dx\,dy\,dz,  \label{axial_width}
\end{align}%
with the results displayed in Fig. \ref{thickness}. As shown in Fig.~\ref%
{thickness}(a), the radial size $\langle \rho\rangle $ increases with the
growth of both $N$ and vorticity $S$. This is attributed to the centrifugal
effect associated with the vorticity, which leads to an outward pressure in
the radial direction, thereby stretching the QD radially. Note that, as
shown in Fig.~\ref{thickness}(a), the radial thickness $\langle \rho \rangle
$ increases with $S$, whereas the peak density $|\psi |_{\max }^{2}$
exhibits a decreasing trend. This behavior is a natural consequence of the
centrifugal effect introduced by the phase winding in the multipole states.
Indeed, as $S$ increases, the system experiences enhanced radial spreading
due to the rotational kinetic-energy pressure, reducing the local density in
the core. This argument highlights the compressible 3D structure of the
multipole QDs, which adjust their shape and density distribution under the
action of the toroidal confinement. The axial thickness $\langle |z|\rangle $%
, plotted in Fig.~\ref{thickness}(b), also increases with $N$, but exhibits
relatively weak variation across different vorticity values. These features
suggest that the axial confinement, being tighter, resists the expansion
caused by the enhanced centrifugal force. Fig.~\ref{s_vs_E} shows the
variation of the chemical potential $\mu $ in panel (a), and energy $E$ in (b)
with respect to $S$ for different atom numbers ($N=1000,2000,3000,5000$). In
Fig.~\ref{s_vs_E}(b), it is observed that the ground state (energy minimum)
corresponds to $S=0$ for each atom number $N$. The energy of the other
configurations increases with $S$ up to $S=10$, while for $S=11$ and $S=12$
both the chemical potential and energy decreases. For a large value of
$S$, the centrifugal term $S^{2}/(2\rho^{2})$
forces the density annulus to expand outward.
Therefore, the droplet makes
both the peak density and LHY contribution lower, leading to the decrease
of $\mu$ and $E$ for $S=11,12$.

\subsection{Dynamical states of multipole droplets under the action of the toroidal confinement}
\begin{figure*}[tbp]
\centering
\xincludegraphics[width=0.4\linewidth, label={$\qquad$a)},pos={n}]{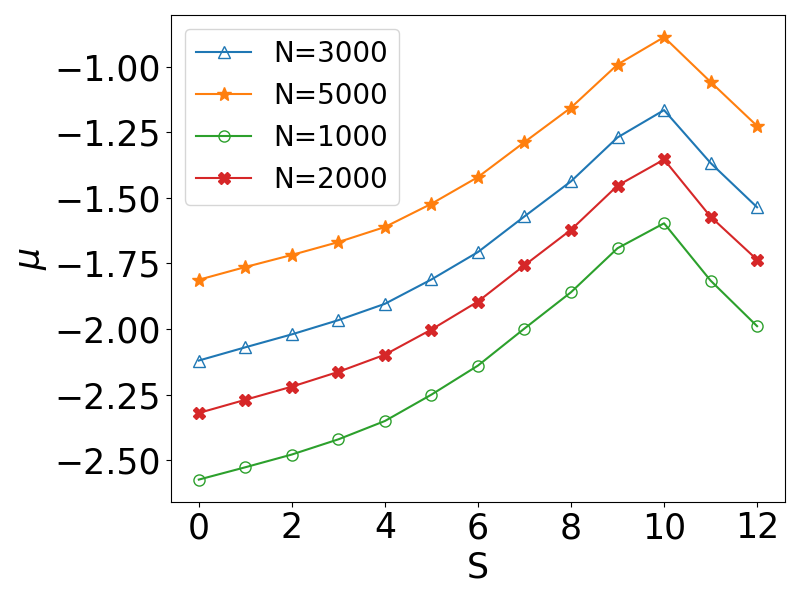} %
\xincludegraphics[width=0.4\linewidth, label={$\qquad$b)},pos={n}]{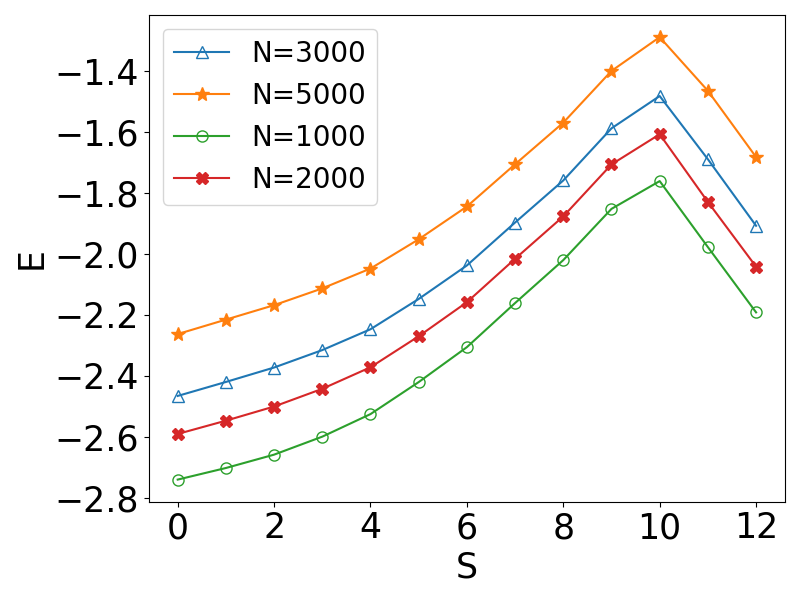}
\caption{The variation of (a) the chemical potential $\protect\mu $ and (b)
total energy $E$ vs. vorticity $S$, as produced by the
stationary solutions for different numbers of atoms with DDI length $a_{%
\mathrm{dd}}=130.8a_{0}$, scattering length $a=100a_{0}$, and $\protect%
\gamma _{\mathrm{QF}}=2.697\times 10^{7}a_{0}^{5/2}$. The ground-state
configuration corresponds to $S=0$, with both the energy $E$ and chemical
potential $\protect\mu $ increasing gradually till $S=10$, followed by the
decrease of $E$ and $\protect\mu $ for $S=11$ and $12$. }
\label{s_vs_E}
\end{figure*}
The above-mentioned anti-VK criterion, $d\mu /dN>0$, observed in Fig.~\ref%
{natoms}(a), is a necessary but not sufficient stability condition. To
assess the full stability of the QDs configurations, we performed the
real-time simulations by perturbing the stationary solutions by $1\%$ random
noise and monitoring the subsequent evolution. In this way, we explicitly
distinguish between the stationary states (Figs. \ref{s_0}-\ref{projection})
and dynamical stability under the action of perturbations. Characteristic
examples of the perturbed dynamics are presented in Fig.~\ref{rl_tme}. It is
observed that the QDs corresponding to $S=0$ remain stable, as shown in Fig.~%
\ref{rl_tme}(a), at time $t=4t_{0}\approx 10.44$ ms, with $t_{0}=2.61\ $ms.
In contrast, the multipolar QDs with $S=1$, $2$, $5$, $6$, $7$, $8$ and $10$
(Figs.~\ref{rl_tme}(b--h)) exhibit clear signatures of instability, with the
instability growth rate increasing with $S$.
The instability times reported here are typically in the range of $1$–$10$ ms,
which is short but still within experimentally relevant timescales
\cite{cidrim2018vortices}. Although they are unstable, QDs persist under small perturbations
with the random noise applied at the level of $0.1\%$ during the
time interval $t=0.4t_{0}\approx 1.044$ ms, as shown in
Fig. \ref{rl_tme_0.1}(a) and (b) for the configurations $S=1$ and $S=6$.
This behavior reflects a transient robustness rather than metastability.
In contrast, the configurations with $S=7$
and $S=8$ undergo fragmentation, as shown in Figs.~\ref{rl_tme_0.1}c)--d).
Figures \ref{rl_tme_lhy}(a) and (b) further explore the role of the LHY correction in the high-vorticity regime at $t=4.0t_{0}\approx
10.44$ ms. For $S=12$, the enhanced centrifugal effect drives neighboring
density lobes to approach each other. When the LHY correction is included,
its additional repulsive contribution counteracts this tendency, preventing
full coalescence and resulting in a partially merged unstable configuration,
as seen in Fig.~\ref{rl_tme_lhy}(a). In contrast, when
$\gamma_{\mathrm{QF}}=0$ (no LHY term), the system enters the attraction
regime that favors the full merger of the lobes, leading to the formation of a
ring-like configuration, displayed in Fig.~\ref{rl_tme_lhy}(b).
\begin{figure}[tbp]
\centering
\xincludegraphics[width=0.37\linewidth, label={a)}]{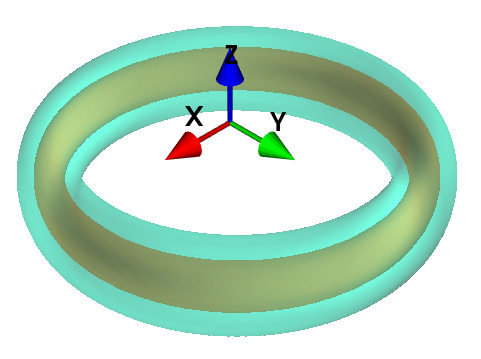} %
\xincludegraphics[width=0.37\linewidth,  label={b)}]{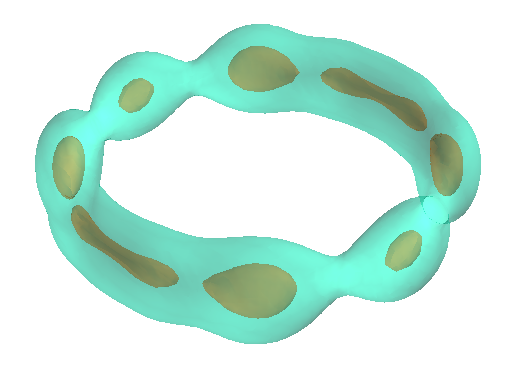} %
\xincludegraphics[width=0.37\linewidth,  label={c)}]{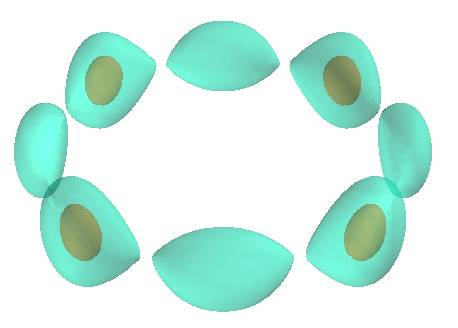} %
\xincludegraphics[width=0.37\linewidth,  label={d)}]{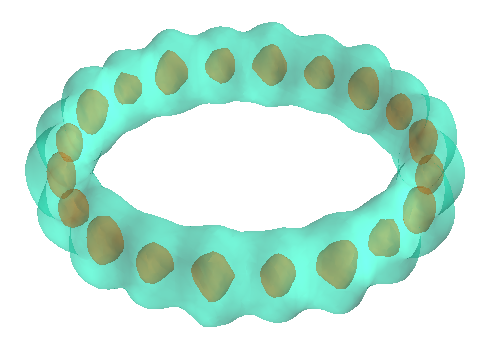} %
\xincludegraphics[width=0.37\linewidth,  label={e)}]{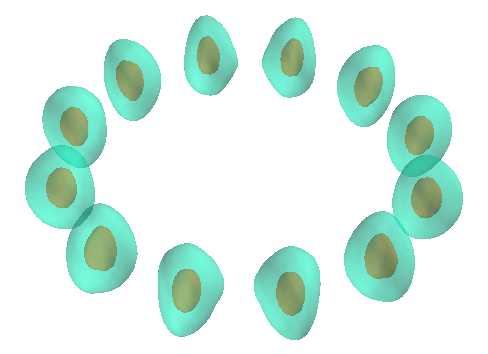}
\xincludegraphics[width=0.37\linewidth,  label={f)}]{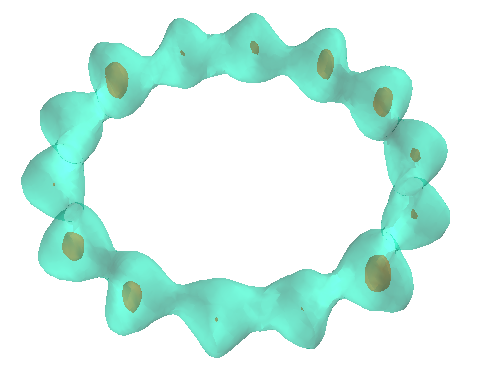}
\xincludegraphics[width=0.37\linewidth,  label={g)}]{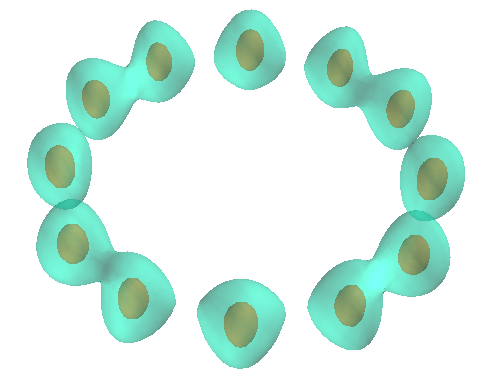}
\xincludegraphics[width=0.37\linewidth,
label={h)}]{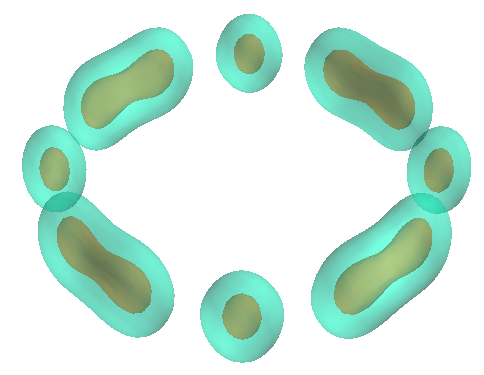}
\caption{The real-time dynamics of QDs perturbed random noise at the $1\%$
level for various vorticities at time $t=4.0t_{0}\approx 10.44$ ms, with $%
t_{0}=2.61$ ms. (a) $S=0$, (b) $S=1$, (c) $S=2$, (d) $S=5$, (e) $S=6$, (f) $%
S=7$, (g) $S=8$, and (h) $S=10$, highlighting the transition from the stable
to unstable behavior under strong perturbations (1\% noise) as $S$
increases. Under the action of this perturbation, only the state with $S=0$
remains stable (black circle in Fig. \protect\ref{stab_boun}).
Higher-vorticity states, with $S\ge1$, exhibit unstable behavior; however,
some of these states persist under the action of weaker perturbations for
short time intervals. All the results are produced for $N=5000
$, $a_{\mathrm{dd}}=130.8a_{0}$, scattering length $a=100a_{0}$, and the LHY
coefficient $\protect\gamma_{\mathrm{QF}}=2.697\times 10^{7}a_{0}^{5/2}$. }
\label{rl_tme}
\end{figure}

To further examine the dynamical behavior of the QDs, we display the perturbed
evolution of their radial $\langle \rho \rangle $ and axial $\langle |z|\rangle $ widths,
as obtained from real-time simulations. To this end, the stationary modes, produced by means of the ITP method,
were perturbed by a $1\%$ random noise and then used as initial conditions for
the real-time propagation, with vorticities $S=0,1,3$ and $N=5000$ atoms.
Figures~\ref{osc}(a) and (b) show the evolution of
$\langle \rho \rangle$ and $\langle |z| \rangle$, respectively,
for trap parameters $d=\sqrt{6}$ and $z_{0}=\sqrt{6}$.
In the case of $S=0$, corresponding to the energetically stable ground state,
the QD remains robust against the applied perturbation, with the radial width
remaining nearly constant and weak oscillations appearing in the axial direction.
In contrast, the vortical configurations with $S=1$ and $S=3$ exhibit pronounced
oscillatory dynamics in both the radial and axial widths, reflecting the excitation
of collective modes by the initial perturbation. For the wider trap with $d=\sqrt{8}$ and $z_{0}=\sqrt{8}$, used in Figs.~\ref{osc}(c,d), the oscillation frequency remains essentially unchanged for all vorticities. However, the oscillation amplitude increases noticeably for the vortical states $S=1$ and $S=3$, as may be expected from trap-induced collective modes.
These observations suggest that the oscillatory dynamics is determined by the external
confinement.

Finally, in Fig.~\ref{stab_boun} we map stability boundaries of QDs with different vorticities $S$, as revealed by systematic real-time simulations of the perturbed evolution. The $S=0$ configuration (black circles) always corresponds to the dynamically stable ground state, preserving its shape over the longest simulated evolution time, $t \approx 10.44$ ms. The configurations with $1 \leq S \leq 6$ (blue stars) remain intact over relatively short evolution times, up to $t \approx 1.044$ ms, reflecting a transient robustness under the action of weak, $0.1\%$, random perturbations. In contrast, states with $S \geq 7$ (up to the largest value considered here, $S=12$, which are denoted by red triangles) are clearly unstable and rapidly decay, even under the action of the weak perturbation ($0.1\%$ ), during the same short
time $t \approx 1.044$ ms. The configurations labeled $8a, 9a-9d, 10a$ along the horizontal line corresponding to $N=5000$ are cross-referenced with the real-time  simulations displayed in Figs. \ref{rl_tme}-\ref{rl_tme_lhy}.
\begin{figure}[tbp]
\centering
\xincludegraphics[width=0.43\linewidth,  label={a)}]{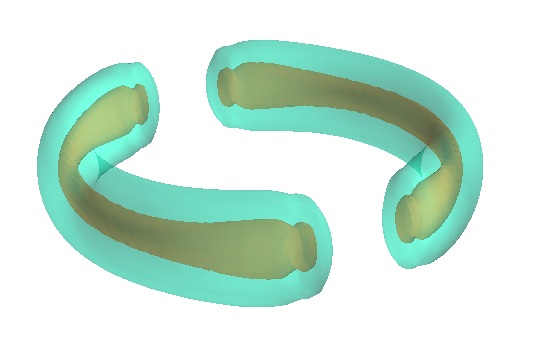} %
\xincludegraphics[width=0.43\linewidth,  label={b)}]{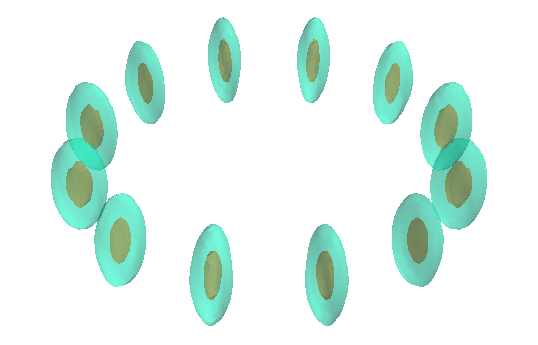} %
\xincludegraphics[width=0.43\linewidth,  label={c)}]{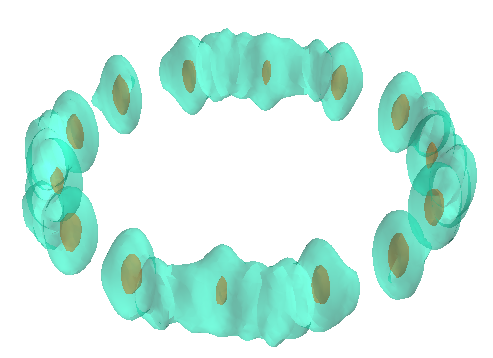} %
\xincludegraphics[width=0.37\linewidth,  label={d)}]{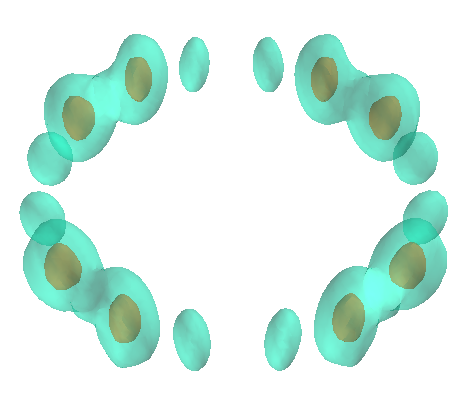}
\caption{The real-time dynamics of QDs perturbed by random noise at the $%
0.1\%$ level, for different vorticities. Panels (a)-(d) display the perturbed
evolution of the QDs for $N=5000$ with the DDI length $a_{\mathrm{dd}%
}=130.8a_{0}$, scattering length $a=100a_{0}$, and the LHY coefficient $%
\protect\gamma _{QF}=2.697\times 10^{7}a_{0}^{5/2}$ at time $%
t=0.4t_{0}\approx 1.044$ ms, with $t_{0}=2.61$ ms. (a) $S=1$, (b) $S=6$, (c) $%
S=7$, (d) $S=8$. The $S=1$ and $S=6$ configurations retain their structure for short times
under weak perturbations, exhibiting only a transient robustness
(represented by blue stars in Fig. \protect\ref{stab_boun}), while the
modes with $S=7$ and $S=8$ are unstable (represented by red triangles in Fig. 
\protect\ref{stab_boun}), decaying even under the action of the weak noise
with the 0.1\% strength.}
\label{rl_tme_0.1}
\end{figure}
\subsection{Stationary states of multipole droplets under Gaussian
Confinement}
Finally, we investigated the effect of the external confinement, replacing
the ring-shaped (toroidal) trap (\ref{toroid}) by the Gaussian potential,
with $\rho _{0}=0$, to analyze the formation and structure of the VQDs under
different geometric constraints. Figure (\ref{shell}) illustrates the
formation of multipole QDs under the action of the Gaussian potential.
Panels (a) - (f) represent 2D density distribution of the stable QD for
$S=1 - 6 $ configurations, respectively. It is observed for higher vorticity values ($S=4,5,6$),
the density distribution exhibits distinct high-density
lobes alternating with low-density regions, reflecting the underlying
angular modulation induced by the higher-order vorticity. The angular
modulation arises due to the inability of the Gaussian trap to directly
accommodate high-vorticity states. Unlike the toroidal trap, which naturally
supports the circulating superflow and allows the density to spread
uniformly along the ring, the centrally peaked Gaussian trapping potential keeps the
condensate near the origin. Therefore, the central confinement and centrifugal effects
lead to the breaking of the axial symmetry and results in  petal-shaped multipole structures.
This comparison demonstrates that multipole droplets also arise from a geometry-dependent balance
between the centrifugal effect, the trapping potential, and the nonlinear interactions.
Thus, the findings of this section reinforce the central result  that trapping geometry plays a
critical role in determining the morphology VQDs, complementing the analysis
presented in Sec.~III.1 and Sec.~III.2.

\begin{figure}[tbp]
\centering
\xincludegraphics[width=0.43%
\linewidth,label={a)}]{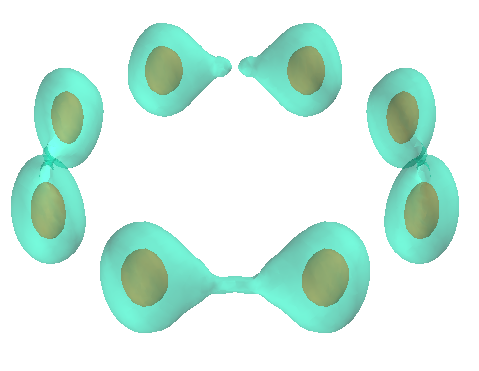} %
\xincludegraphics[width=0.47\linewidth,  label={b)}]{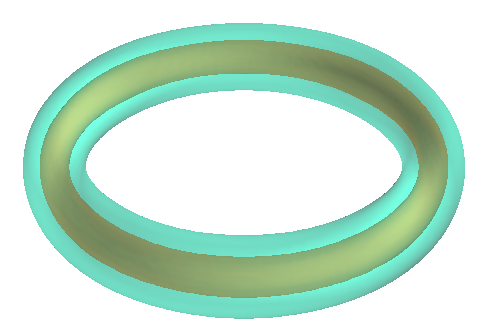}
\caption{The role of the LHY correction is illustrated by the real-time
dynamics of QDs with vorticity $S=12$ and atom number $N=5000$, DDI
length $a_{\mathrm{dd}}=130.8a_{0}$, and scattering length $a=100a_{0}$,
evaluated at $t=4.0t_{0}\approx 10.44$ ms with $t_{0}=2.61$ ms.
Panel (a) includes the LHY correction term with $\protect\gamma
_{\mathrm{QF}}=2.697\times 10^{7}a_{0}^{5/2}$. Due to enhanced centrifugal
effect, the poles begin to merge with the neighboring poles, while the
repulsive LHY term prevents the complete merging, resulting in an unstable
configuration (represented by a red triangle in Fig. \ref{stab_boun}). Panel (b) corresponds to the case without the correction, $\protect\gamma_{\mathrm{QF}}=0$, resulting in an attractive regime leading to
complete merger of the droplets into a ring-shaped configuration. All results are obtained in the absence of external perturbations.}
\label{rl_tme_lhy}
\end{figure}
\section{Conclusion}
We have investigated the emergence, structure, and stability of the 3D
ring-shaped and multipole QDs (quantum droplets) in the dipolar BEC confined
by the toroidal potential. The interplay of the DDI (dipole-dipole
interactions) with the beyond-mean-field LHY (Lee-Huang-Yang) correction,
produces a broad variety of the vorticity-carrying 3D QD necklace-like
arrays.
\begin{figure}[tbp]
\centering
\xincludegraphics[width=0.490\linewidth, label={$\quad$a)},pos={n}]{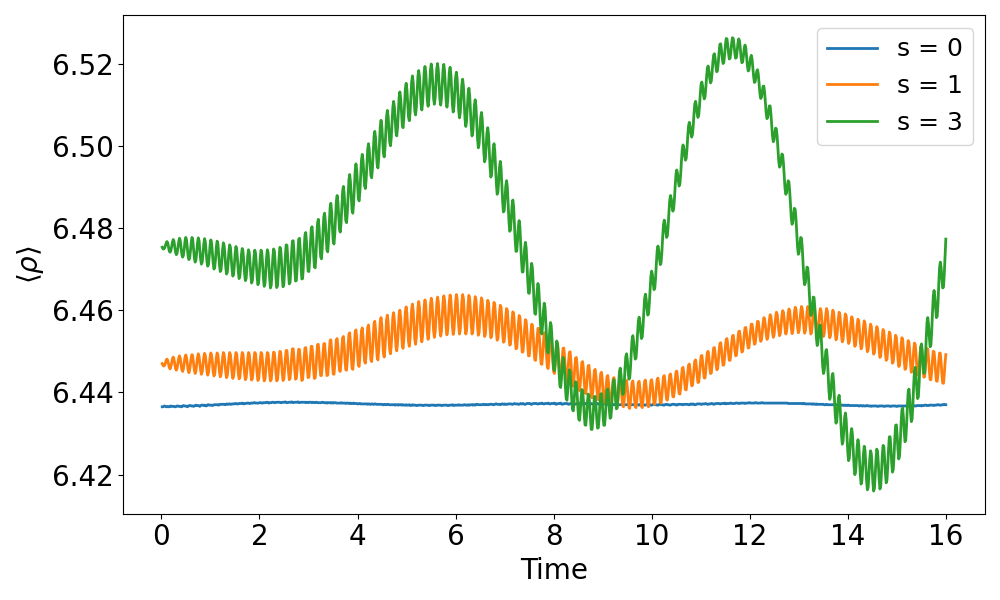} %
\xincludegraphics[width=0.490\linewidth, label={$\quad$b)},pos={n}]{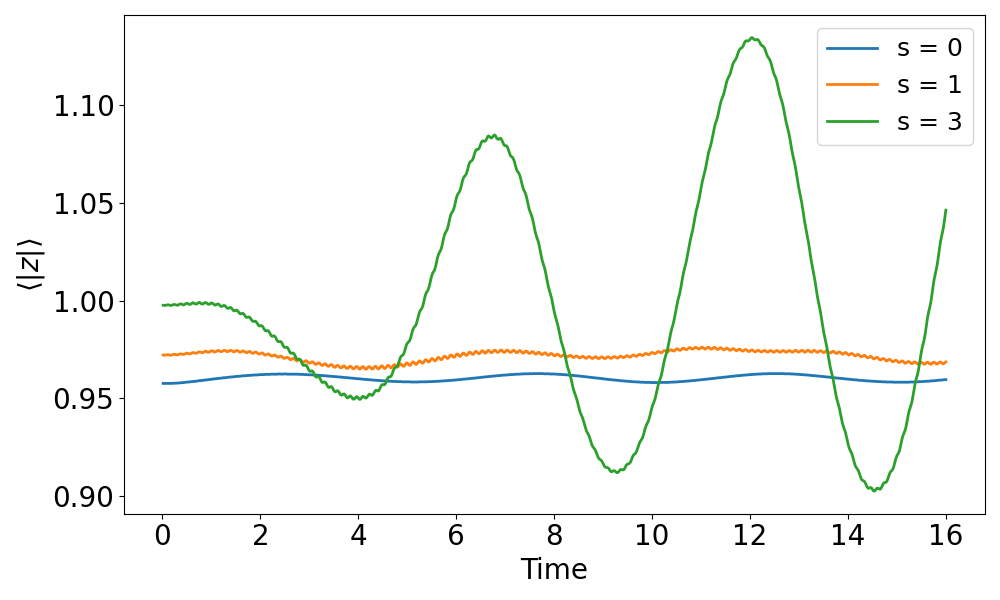}
\xincludegraphics[width=0.490\linewidth, label={$\quad$c)},pos={n}]{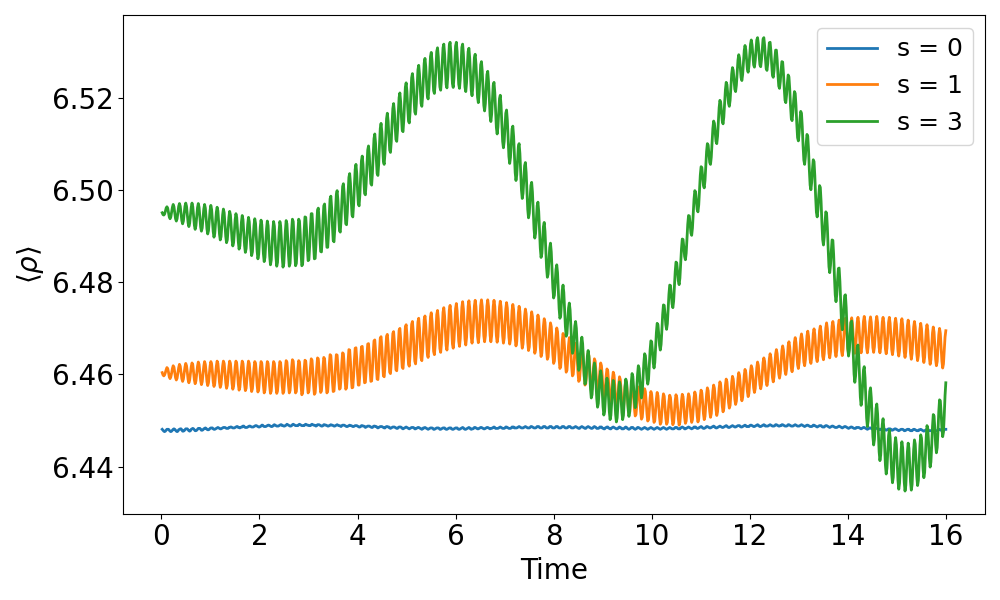} %
\xincludegraphics[width=0.490\linewidth, label={$\quad$d)},pos={n}]{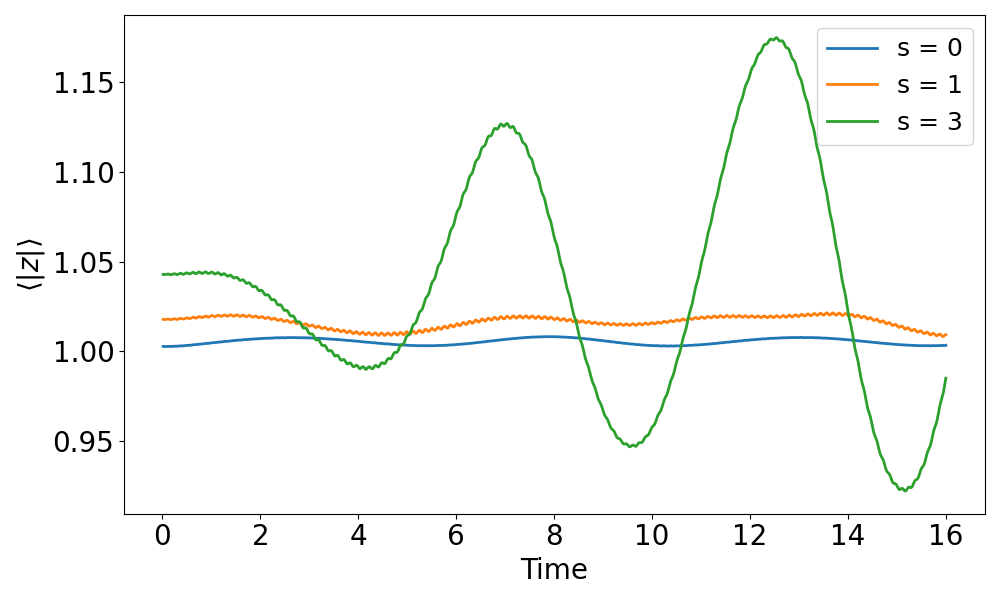}
\caption{Real-time evolution of the radial  $\langle \rho \rangle$ (a,c)
and axial $\langle |z| \rangle$ (b,d) widths of QDs with vorticities
$S=0,1,$ and $3$, $N=5000$ atoms, DDI length
$a_{\mathrm{dd}}=130.8a_{0}$, scattering length $a=100a_{0}$, and
$\gamma_{\mathrm{QF}}=2.697\times10^{7}a_{0}^{5/2}$, for different trap widths.
Panels (a,b) correspond to the trap parameters $d=\sqrt{6}$ and
$z_{0}=\sqrt{6}$ with $p=6$ and $\rho_{0}=2\pi$, while panels (c,d) correspond to a wider trap with
$d=\sqrt{8}$ and $z_{0}=\sqrt{8}$ and the same values $p=6$ and $\rho_{0}=2\pi$.
The stationary state, produced by means of ITP for each vorticity, is used as the initial condition
for real-time evolution, which is perturbed by a $1\%$ random noise.
For panels (a,b), the $S=0$ droplet remains stable, exhibiting an
almost constant radial width and weak axial oscillations, whereas the
vortical configurations with $S=1$ and $S=3$ display growing oscillatory
dynamics in both radial and axial directions. For the wider trap in panels (c,d),
the oscillation frequency remains unchanged, while the oscillation amplitude
increases for all the states, indicating enhanced excitation of the
trap-induced collective modes.
}
\label{osc}
\end{figure}
\begin{figure}[tbp]
\centering
\xincludegraphics[width=1\linewidth]{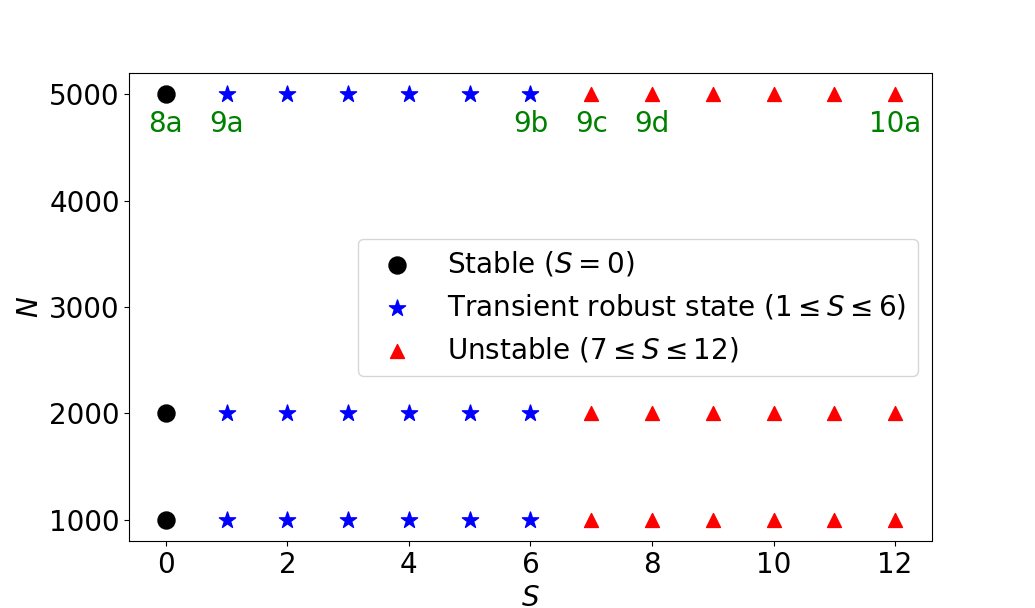}
\caption{Stability boundaries of the QD ring-shaped configurations, as
inferred from the systematic simulations of the perturbed evolution, for
different values of the atom number $N$ and vorticity $S$. Black circles represent stable ground-state configurations obtained for $S=0$, which preserve their structure up to $t=4.0\,t_{0}\approx10.44$ ms under the action of the $1\%$ random perturbation. States with $1\leq S\leq6$, indicated by blue stars, are short-lived (transiently stable) ones, retaining their structure only up to $t=0.4\,t_{0}\approx1.044$ ms, under the action of a weaker $0.1\%$ perturbation. Configurations with $S\geq7$ (up to the largest value considered here, $S=12$), denoted by red triangles, are unstable, decaying under the action of the weak $0.1\%$ perturbation during the time $t\approx1.044$ ms. The markers placed at the horizontal line corresponding to the atom number $N=5000$ refer to the simulation results presented above:
the point labeled $8a$ (the black circle) corresponds to Fig.~\ref{rl_tme}(a), the points labeled $9a$–$9d$ (blue stars) correspond to Figs.~\ref{rl_tme_0.1}(a–d),
and the point labeled $10a$ (the red triangle) corresponds to Fig.~\ref{rl_tme_lhy}(a). The other parameters are $a_{\mathrm{dd}}=130.8a_{0}$, $a=100a_{0}$,
and $\gamma _{\mathrm{QF}}=2.697\times 10^{7}a_{0}^{5/2}$.}
\label{stab_boun}
\end{figure}
\begin{figure}[tbp]
\centering
\xincludegraphics[width=0.48\linewidth,label={a)}]{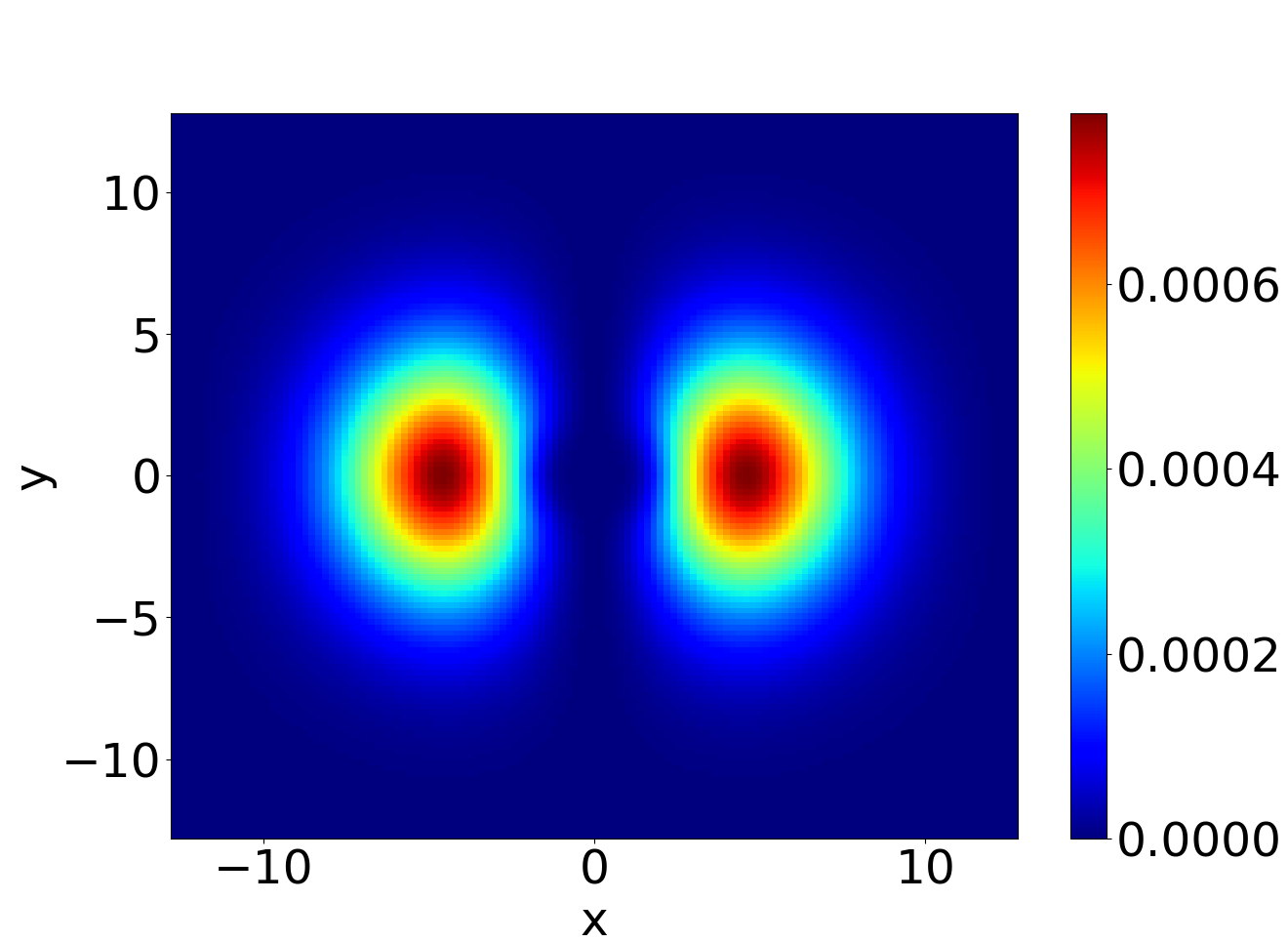}
\xincludegraphics[width=0.48\linewidth,label={b)}]{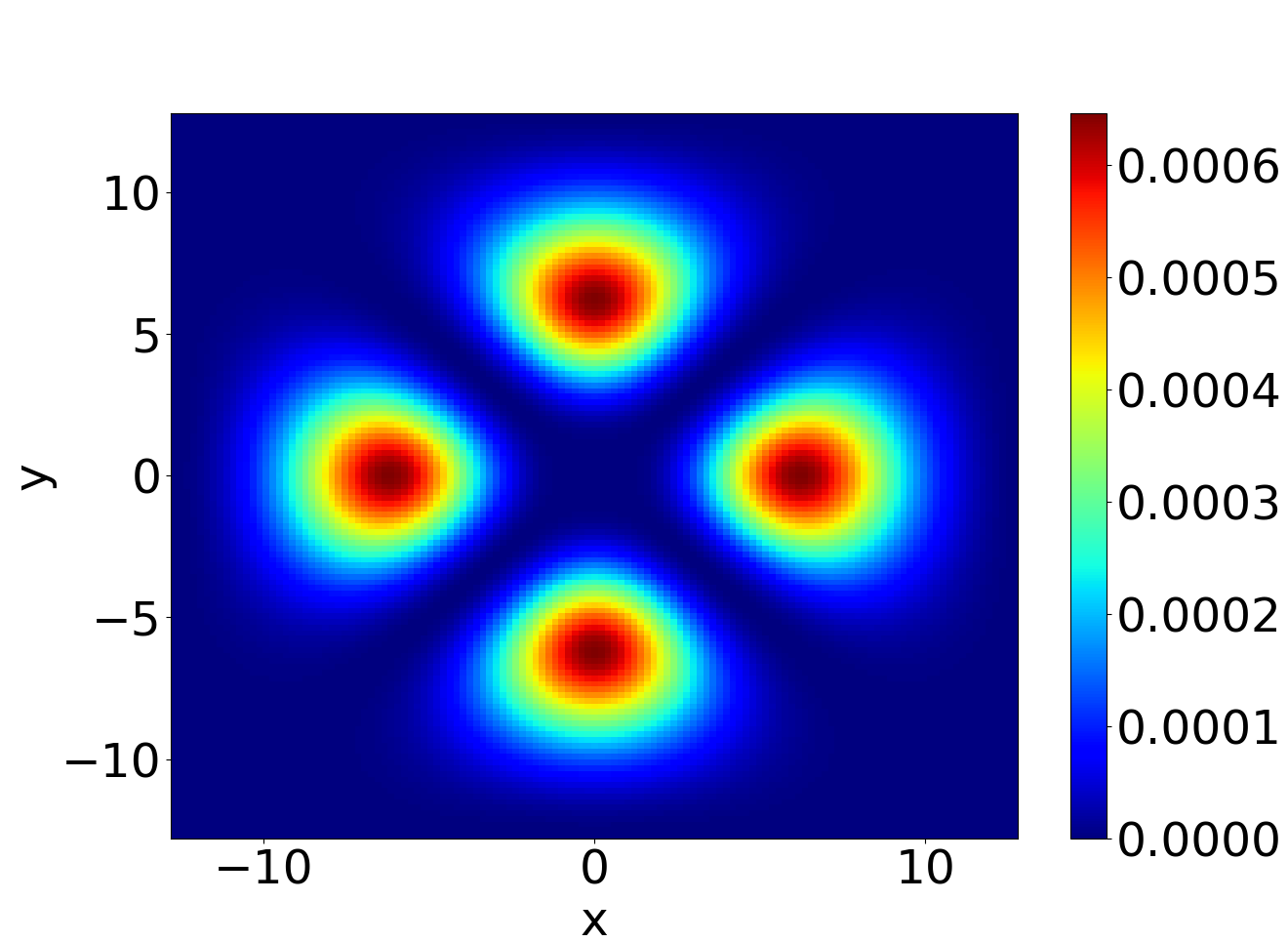}
\xincludegraphics[width=0.48\linewidth,label={c)}]{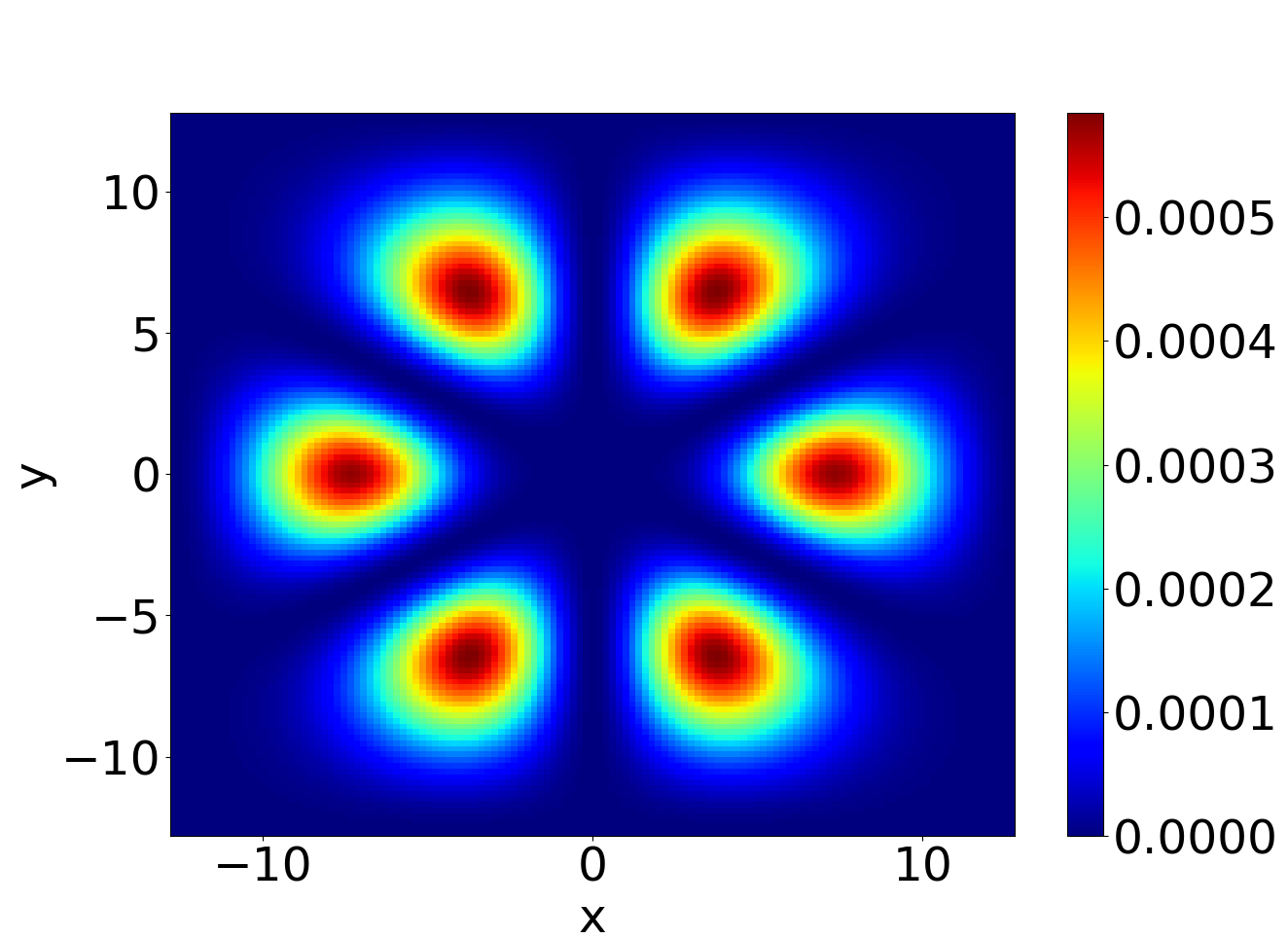}
\xincludegraphics[width=0.48\linewidth,label={d)}]{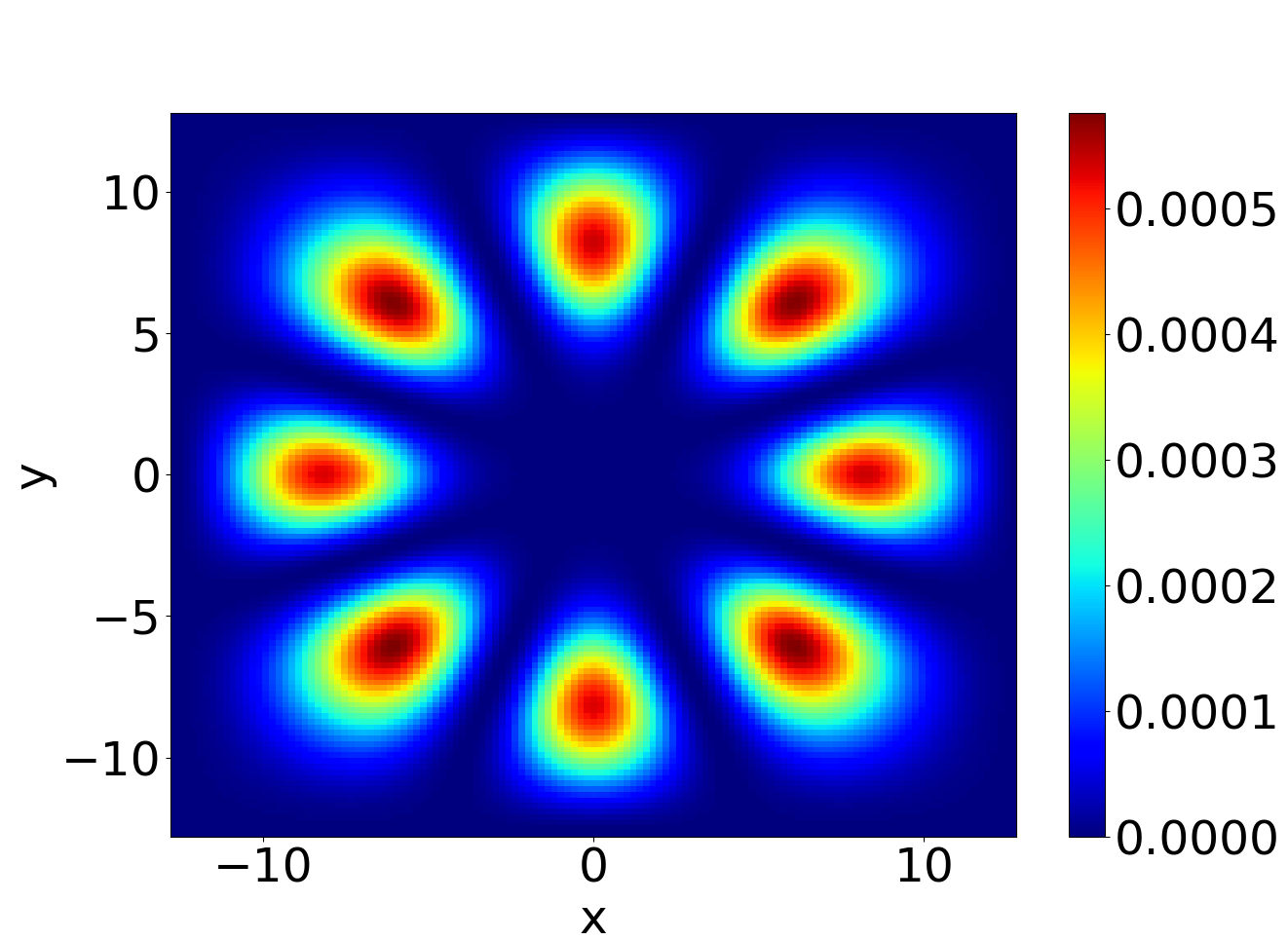}
\xincludegraphics[width=0.48\linewidth,label={e)}]{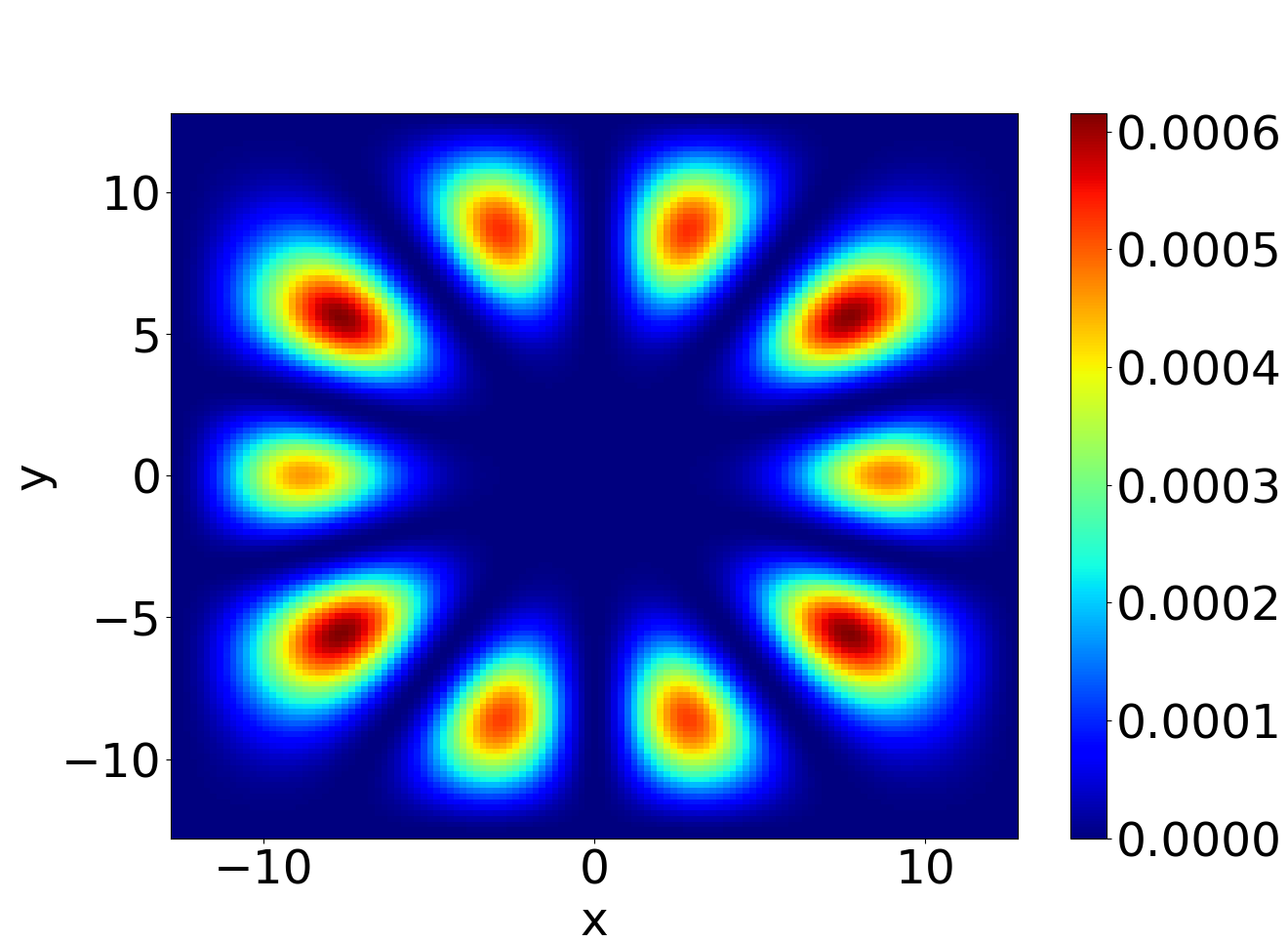}
\xincludegraphics[width=0.48\linewidth,label={f)}]{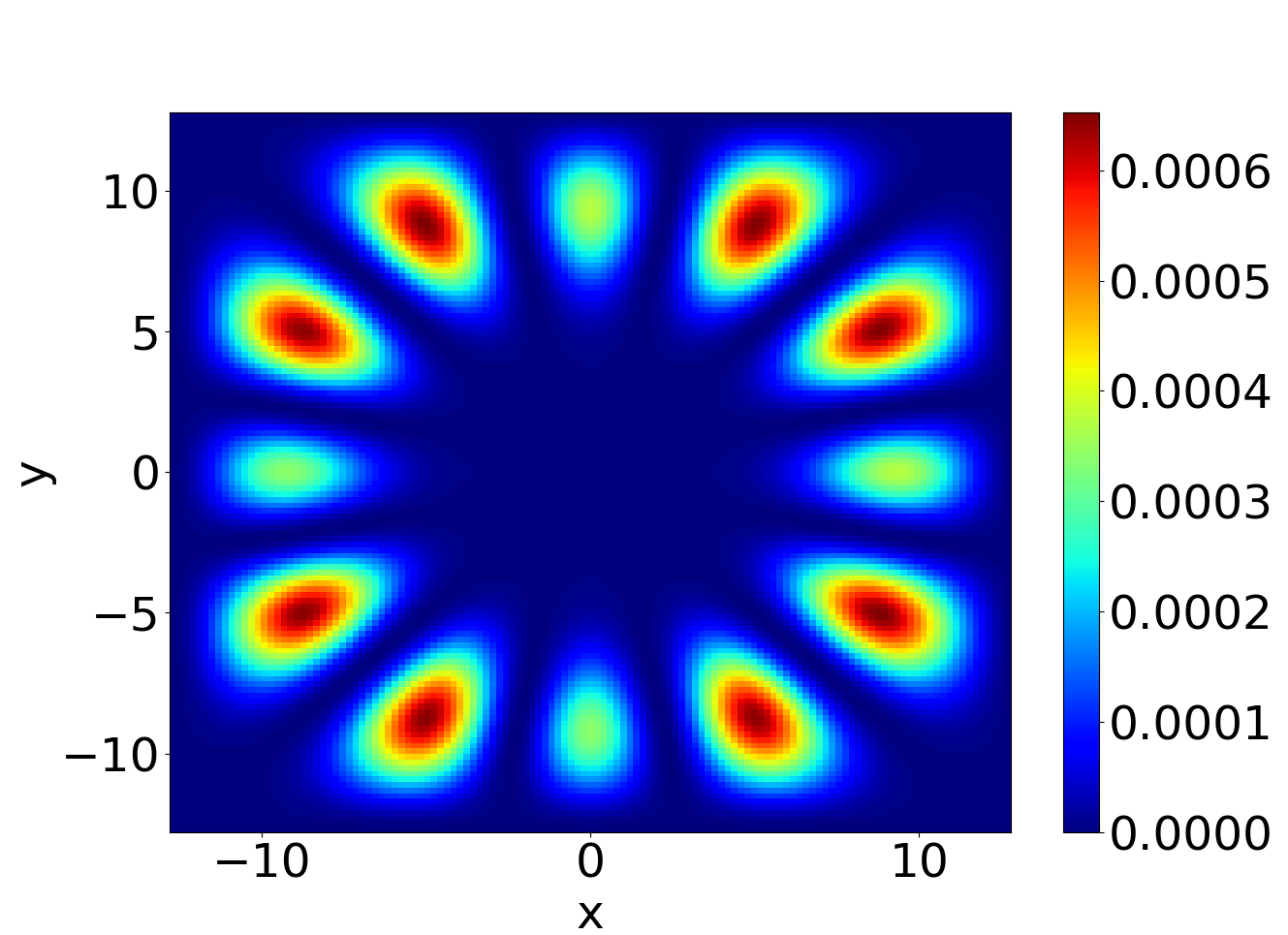}
\caption{Contour plots of the 2D density projection
of stable multipole QDs in the $(x,y)$ plane, under the action of the
Gaussian confinement represented by potential (\protect\ref{toroid}) with
$\rho_{0}=0$, $p=-10$, $d=z_{0}=1$. Panels (a) - (f) pertain to
vorticities $S=1-6$, respectively. An increase in the vorticity
results in structural variations of the QDs. All the
results are obtained for the atom number
$N=2000$, DDI length $a_{\mathrm{dd}}=130.8a_{0}$, scattering length $%
a=100a_{0}$, and the strength of the LHY correction $\protect\gamma _{\mathrm{QF}%
}=2.697\times 10^{7}a_{0}^{5/2}$.}
\label{shell}
\end{figure}
Our results demonstrate that the toroidal geometry naturally facilitates the
formation of VQDs (vortex QDs), which exhibit donut-shaped density
distributions with the internal phase-imprinted vorticity. With increasing
vorticity $S$, the VQD configurations evolve into multipole patterns, with
the number of density peaks (poles) approximately given by $n=2S$, up to $%
S\leq 6$. The multipolar structures are stabilized by the combined action of
the confinement geometry, nonlocal DDI, and the effect of quantum
fluctuations represented by the LHY term. Remarkably, even in the absence of
the latter one, the QD remains robustly self-bound, as the DDI alone can
support stable states in the combination with the trapping potential,
preventing the onset of the collapse.
Increasing the atom number $N$, naturally leads to an expansion of the
droplet along the azimuthal direction. Through systematic numerical
simulations, we have investigated the effects of varying $N$ and vorticity $%
S $ on the chemical potential, energy, amplitude, and radial and axial
widths of the QDs. They satisfy the anti-VK necessary stability condition,
where the full stability has been verified by means of systematic simulations
of the perturbed evolution. The total energy of the QD stationary states
gradually increases as a function of $N$, higher-vorticity states exhibiting
larger energies due to the centrifugal effect, which also leads to the
reduction in the peak density as the radial expansion of the QD structure.
The radial width consistently increases with $N$ for all values of the
vorticity, as expected. In contrast, the axial ($z$) width shows only minor
variations, indicating an essentially tighter confinement in that direction.

The stability analysis based in the direct simulations demonstrates that,
while the ground-state QDs with $S=0$ exhibit complete stability, multipole
VQDs became increasingly susceptible to structural deformations and,
eventually, fragmentation with the increase of vorticity $S$. The results
are summarized as follows: the configurations with $1\leq S\leq 6$
survive under the action of mild random perturbations, with the relative
amplitude $\sim $ $0.1\%$, whereas a stronger perturbation with the amplitude $\sim 1\%$
leads to fragmentation, indicating the transient robustness.
In contrast, the ones with $S\geq 7\ $are definitely unstable.
We have also highlighted the significant role of the trapping geometry, by
comparing the properties of the QDs in the toroidal and Gaussian
confinements. The toroidal potential supports persistent superflow, allowing
the stable higher-order states. However, when the trap is replaced by the
centrally peaked Gaussian, the condensate undergoes symmetry breaking,
resulting in angular fragmentation even at moderate vorticities. This
outcome originates from the interplay of the azimuthal dynamics and trapping
geometry, with the competition of the centrifugal expansion and the central
localization leading to the spontaneous pattern formation.
In summary, this work presents a comprehensive analysis of the 3D
ring-shaped and multipole QDs in the dipolar BEC under the combined action
of the toroidal trap and LHY correction. These findings advance the
theoretical understanding of structured states of the quantum matter and
offer insights in the possibility of the experimental realization of novel
VQD states in ultracold gases of magnetic atoms, such as $^{164}$Dy. Future
directions of the studies may be aimed at the role of spin-orbit coupling
and overall rotation, potentially uncovering new regimes of supersolidity,
topological excitations, and quantum turbulence in dipolar condensates.
\newline
\newline
\textbf{Acknowledgment:} S. Sanjay and S. Saravana Veni acknowledge Amrita
Vishwa Vidyapeetham, Coimbatore, where this work was supported under Amrita
Seed Grant (File Number: ASG2022141).
\end{subequations}

\end{document}